\documentclass[twocolumn,nolinenumbers]{aastex631}
\usepackage{graphicx}
\usepackage{epstopdf}
\usepackage{subfigure} 
\usepackage{color}
\usepackage{float}
\usepackage[figuresright]{rotating}
\usepackage{footnote}
\usepackage{amsmath}
\usepackage{booktabs}
\usepackage{threeparttable}
\usepackage{hyperref}

\shorttitle{The High-precision Empirical Stellar Mass Library}
\shortauthors{J.P Xiong et al.}
\graphicspath{{./}{figures/}}

\usepackage{enumitem}
\setenumerate[1]{itemsep=0pt,partopsep=0pt,parsep=\parskip,topsep=5pt}
\setitemize[1]{itemsep=0pt,partopsep=0pt,parsep=\parskip,topsep=5pt}
\setdescription{itemsep=0pt,partopsep=0pt,parsep=\parskip,topsep=5pt}

\newcommand{\kms}{km\ s$^{-1}$}

\newcommand{\teff}{$T_{\rm eff}$}
\newcommand{\logg}{$\log g$}
\newcommand{\mh}{[M/H]}
\newcommand{\mass}{$M$}
\newcommand{\radius}{$R$}

\newcommand{\lum}{$L$}
\newcommand{\masssol}{$M_\odot$}

\def\tabnotefont{\fontsize{8}{9}\selectfont}
\def\x{@{\extracolsep{\fill}}}
\def\Hline{%
  \noalign{\ifnum0=`}\fi\hrule \@height 1.5pt \futurelet
   \@tempa\@xhline}
\newenvironment{tabnote}{\par\vskip1.5pt%
  \tabnotefont
}{%
  \par}

\begin{document}
\nolinenumbers
\title{The Eclipsing Binaries from the LAMOST Medium-resolution Survey.III. A High-precision Empirical Stellar Mass Library}
\correspondingauthor{Chao Liu}
\email{liuchao@nao.cas.cn}

\author[0000-0003-4829-6245]{Jianping Xiong}
\affiliation{Key Laboratory of Optical Astronomy, National Astronomical Observatories, Chinese Academy of Sciences\\
Beijing 100101, China}
\affiliation{University of Chinese Academy of Sciences\\
Beijing 100049, China}

\author[0000-0002-1802-6917]{Chao Liu$^{\dag}$}
\affiliation{Key Laboratory of Space Astronomy and Technology, National Astronomical Observatories, Chinese Academy of Sciences\\
Beijing 100101, China}
\affiliation{Institute for Frontiers in Astronomy and Astrophysics, Beijing Normal University, Beijing, 102206, China}
\affiliation{University of Chinese Academy of Sciences\\
Beijing 100049, China}

\author[0000-0002-2577-1990]{Jiao Li}
\affiliation{Key Laboratory of Space Astronomy and Technology, National Astronomical Observatories, Chinese Academy of Sciences\\
Beijing 100101, China}

\author[0000-0002-3651-5482]{Jiadong Li}
\affiliation{Key Laboratory of Space Astronomy and Technology, National Astronomical Observatories, Chinese Academy of Sciences\\
Beijing 100101, China}
\affiliation{University of Chinese Academy of Sciences\\
Beijing 100049, China}

\author[0000-0002-6434-7201]{Bo Zhang}
\affiliation{Key Laboratory of Space Astronomy and Technology, National Astronomical Observatories, Chinese Academy of Sciences\\
Beijing 100101, China}

\author[0000-0001-7084-0484]{Xiaodian Chen}
\affiliation{Key Laboratory of Optical Astronomy, National Astronomical Observatories, Chinese Academy of Sciences,\\ Beijing 100101, China}
\affiliation{School of Astronomy and Space Science, University of the Chinese Academy of Sciences, Beijing 101408, China} 
\affiliation{Department of Astronomy, China West Normal University, Nanchong, China}

\author{Changqing Luo}
\affiliation{Key Laboratory of Space Astronomy and Technology, National Astronomical Observatories, Chinese Academy of Sciences\\
Beijing 100101, China}

\author{Zihuang Cao}
\affiliation{Key Laboratory of Optical Astronomy, National Astronomical Observatories, Chinese Academy of Sciences\\
Beijing 100101, China}
\affiliation{School of Astronomy and Space Science, University of the Chinese Academy of Sciences, Beijing 101408, China}

\author{Yongheng Zhao}
\affiliation{Key Laboratory of Optical Astronomy, National Astronomical Observatories, Chinese Academy of Sciences\\
Beijing 100101, China}
\affiliation{School of Astronomy and Space Science, University of the Chinese Academy of Sciences, Beijing 101408, China}

\begin{abstract}
High-precision stellar mass and radius measured directly from binaries can effectively calibrate the stellar models. However, such a database containing full spectral types and large range of metallicity is still not fully established. A continuous effort of data collecting and analysis are requested to complete the database. In this work, we provide a catalog containing 184 binaries with independent atmospheric parameters and accurate masses and radii as the benchmark of stellar mass and radius. The catalog contains 56 new detached binaries from LAMOST Medium-resolution spectroscopic (MRS) survey and 128 detached eclipsing binaries compiled from previous studies. We obtain the orbital solutions of the new detached binaries with uncertainties of masses and radii smaller than 5\%. These new samples densify the distribution of metallicity of the high-precision stellar mass library and add 9 hot stars with \teff $>8000$ K. Comparisons show that these samples well agree with the PARSEC isochrones in \teff-\logg-mass-radius-luminosity space. We compare mass and radius estimates from isochrone and SED fitting, respectively, with those from the binary orbital solution. We find that the precision of the stellar-model dependent mass estimates is $>10$\% and the precision of the radius estimates based on atmospheric parameters is $>15$\%. These give a general view of the uncertainty of the usual approaches to estimate stellar mass and radius. 
\end{abstract}


\section{Introduction} \label{sec:intro}

Stars, as the main ingredient of galaxies, play critical role in the structure and evolution of galaxies. An accurate understanding of their mass, radius, luminosity, chemical composition, and age is therefore an essential task. In general, The stellar mass and radius can be mainly derived from the stellar structure and evolution models, e.g. PARSEC  \citep{2012MNRAS.427..127B}, $Y^2$ \citep{2004ApJS..155..667D,2008Ap&SS.316...31D}, Dartmounth \citep{2008ApJS..178...89D}, MIST \citep{2016ApJS..222....8D} etc., empirical relations, e.g. mass-luminosity relation (MLR), mass-radius relation (MRR), and mass-temperature relation (MTR) \citep{2014PASA...31...24E,2015AJ....149..131E,2018MNRAS.479.5491E}, or asteroseismic techniques \citep{2014ApJS..210...20M,2014ApJS..210....1C}. 
Among these methods, dynamical mass of the detached eclipsing binaries is not only the method that minimally relies on stellar models, but also reaches precision of around 1\%~\citep{2010AARv..18...67T}. Therefore, it can be used as calibrator for the other mass estimation approaches. 
Indeed, most of the stellar models and empirical relations rely on the dynamical masses and radii for calibration~\citep{2000asqu.book.....C,1973A&A....26..437H,2010AARv..18...67T,2015ASPC..496..164S,2015AJ....149..131E,2018MNRAS.479.5491E}. 

However, by now, only $\sim700$ binaries are provided with accurate masses and radii. Among them, only $<200$ binary systems have the comprehensive atmospheric parameters (\teff, \logg, [M/H]). 
Metallicity is the especially important parameter accompany with stellar mass and radius, since it can help to break the mass-metallicity degeneracy at around turn-off point \citep{2021A&ARv..29....4S}.
With metallicity, one can derive the luminosity-mass and other relations for different metallicity so that the empirical mass can be well compared with stellar models.

It is clear that enlarging the dataset of stars with accurate stellar mass and metallicity can significantly improve the calibration of the stellar models. Recently, significant progress has been made in spectroscopic surveys, e.g. RAVE \citep{2006AJ....132.1645S, 2020AJ....160...83S}, SDSS/SEGUE \citep{2009AJ....137.4377Y}, LAMOST \citep{2012RAA....12.1197C, 2012RAA....12..735D, 2012RAA....12..723Z, 2015RAA....15.1095L}, APOGEE \citep{2017AJ....154...94M}, and GALAH \citep{2015MNRAS.449.2604D} and time-domain photometric surveys, e.g. Gaia \citep{2012Msngr.147...25G, 2004MNRAS.354.1223K, 2018A&A...616A...5C}, NASA/Kepler \citep{2010Sci...327..977B,2010ApJ...713L.109B,2016AJ....151..101A}, TESS \citep{2015JATIS...1a4003R}, ZTF \citep{2019PASP131a8002B}, and ASAS-SN \citep{2017PASP..129j4502K,2019MNRAS.485..961J}. 

In particular, LAMOST has launched a medium-resolution survey (MRS, R$\sim$7500), which includes both time-domain and usual spectroscopic observations~\citep{2020arXiv200507210L}, since October 2018. Moreover, relatively high-quality parameters from the spectra have been obtained. For LAMOST low-resolution spectra (LRS with R$\sim$1800), the accuracy of radial velocity, \teff, \logg \,and [Fe/H] are around 5\,\kms, 150\,K, 0.25\,dex, and 0.15\,dex, respectively~\citep{2015MNRAS.448..822X}. For LAMOST MRS, the accuracy of radial velocity can reach around 1\,\kms\ \citep{2019ApJS..244...27W,2021ApJS..256...14Z,2021RAA....21..265X}. The accuracy of \teff, \logg, and [Fe/H] are around 119\,K, 0.17\,dex, and 0.06$\sim$0.12\,dex, respectively \citep{2020ApJ...891...23W}. Furthermore, many eclipsing binaries with light curves \citep{2011AJ....142..160S,2021MNRAS.503..200J,2020ApJS..249...18C} are identified from LAMOST survey~\citep{2021ApJS..256...31L,2022ApJS..258...26Z}. These published data offer a unique opportunity to measure the dynamic mass and radius of more eclipsing binaries. 

This work aims to present a high-precision empirical stellar mass library that includes the accurate dynamical mass and the independently measured atmospheric parameters (\teff, \logg, [M/H]) from observed spectra for main-sequence stars. The literature data compilation and LAMOST data acquisition and light curve are described in Section~\ref{sec:data}. The method of measuring accurate mass for LAMOST binaries is described in Section~\ref{sec:method}. The results are indicated in Section~\ref{sec:result}. The comparison of dynamical mass and radius derived from different measurement approaches is discussed in Section~\ref{sec:Discussion}. Finally, we summarize in Section~\ref{sec:Conclusion}.

\section{Data} \label{sec:data}
\subsection{Literature data compilation}\label{subsect:literature}

Although previous studies have already published $\sim$700 binary systems with accurate stellar mass and radius~\citep{2010AARv..18...67T,2015ASPC..496..164S,2015AJ....149..131E,2018MNRAS.479.5491E,2021AJ....161..172D}, we can only make use of $<200$ samples with comprehensive atmospheric parameters (\teff, \logg, [M/H]) among the full dataset to provide masses and radii in different metallicity. We select the samples from literature under the following criteria: 1)
the binaries are composed of two main-sequence companions; 2) the masses and radii are estimated from binary orbital dynamics, and 3) \teff, \logg, and [M/H] of (at least) the primary stars are independently derived from spectroscopic data. We finally obtained 128 binaries, 19 from \citet{2010AARv..18...67T}, 78 from \citet{2015ASPC..496..164S}, 24 from \citet{2018MNRAS.479.5491E}, 5 from \citet{2021AJ....161..172D}, and the remaining 2 from \citet{2021MNRAS.504.4302W} and \citet{Pan_2020}, respectively. These data have the uncertainties of the mass and radius estimates at around 1$\sim$2\%. Their metallicity precision is as high as about 0.05\,dex.

These samples are still not sufficient if we want to demonstrate the mass-luminosity relation and other important relations in a large range of effective temperature at different metallicity. Therefore, we need to extend the sample based on the LAMOST survey data.

\subsection{LAMOST data with light curves}\label{subsect:lamostdata}

LAMOST is a 4-meter class reflective Schmidt telescope with a 5-degree field-of-view. Totally, 4000 fibers are installed at its 1.75\,m-diameter focal plane. These fibers go into 16 spectrographs, each of which accepts 250 fibers so that it can take the spectra of 4,000 targets simultaneously. At resolution of $R\sim1800$, LAMOST reaches about $r\sim18$\,mag with 1.5-hour exposure. As October 2018, LAMOST has started the 5-year medium-resolution spectroscopic (MRS) survey ($R\sim7500$ with limiting magnitude of $G<15$\,mag), which includes a time-domain spectroscopic survey sub-project. It is expected that, after a 5-year MRS survey, more than a hundred thousands stars will be observed with $\sim$60 exposures \citep{2020arXiv200507210L}. This provides a unique opportunity to collect a larger sample of binary stars with their orbits resolved. 

The uncertainty of radial velocities of LAMOST MRS spectra is around 1\,\kms\ for late-type stars \citep{2019ApJS..244...27W,2021ApJS..256...14Z,2021RAA....21..265X}, the stellar atmospheric parameters (\teff\ and \logg) and 13 chemical abundances are derived for LAMOST MRS by a deep-learning method with uncertainties of 119\,K, 0.17\,dex, and 0.06$\sim$0.12\,dex for \teff, \logg, and elemental abundances, respectively \citep{2020ApJ...891...23W}. As expected, the rotation of the stars can also be determined under uncertainty of 10\,\kms. For LAMOST LRS, LSP3 can achieve an accuracy of 5\,\kms, 150 K, 0.25 dex, and 0.15 dex for radial velocity, \teff, \logg, and [Fe/H], respectively~\citep{2015MNRAS.448..822X}.

We initially identified 1502 EA-type eclipsing binaries with both LAMOST DR8 MRS multi-epoch spectra and light curves from Kepler eclipsing binary catalog \citep{2011AJ....141...83P}, ZTF Data Release 2 \citep{2020ApJS..249...18C}, and ASAS-SN catalog \citep{2020MNRAS.491...13J}. Then the following criteria are used to select high quality samples: 
\begin{enumerate}
    \item[1)]the double-line spectroscopic binaries are selected;
    \item[2)]the binaries with at least 30 exposures with signal-to-noise ratio larger than 10 are selected;
    \item[3)]atmospheric parameters of primary stars have been measured at the time of secondary minimum (when the primary star obscures the secondary star).
\end{enumerate}
Finally, we select 56 systems from LAMOST DR8 MRS and further estimate the masses and radii by resolving their orbital solutions. Their radial velocity velocity curves were derived by \citet{2021ApJS..256...14Z} via LAMOST multiple-epoch spectra.

\begin{figure*}[t]
  \centering
   \includegraphics[height=22cm, width=19cm]{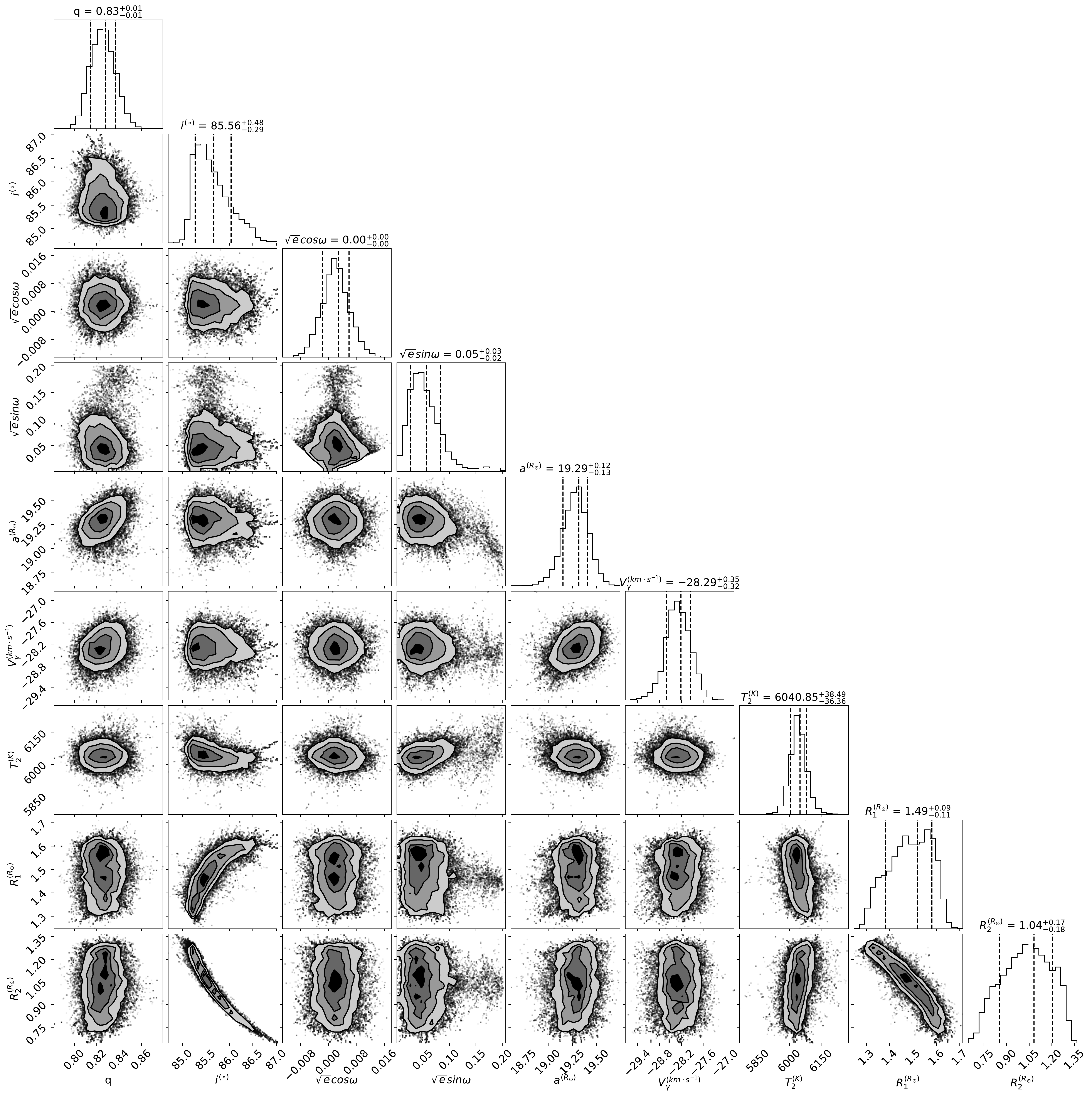}
  \caption{As a sample of the orbital solution, this figure shows the MCMC derived the probability density function of the orbital parameters for KIC 5359678 \citep{2021MNRAS.504.4302W}. From left column to the right, the histograms are for $q$, $i$, $\sqrt{e}cos\omega$, $\sqrt{e}sin\omega$, $a$, $V_{\gamma}$, $T_{2}$, $R_{1}$, and $R_{2}$.}\label{fig:1}
\end{figure*} 

\begin{figure*}[t]
  \centering
  \subfigure[]{
   \includegraphics[height=6cm, width=7cm]{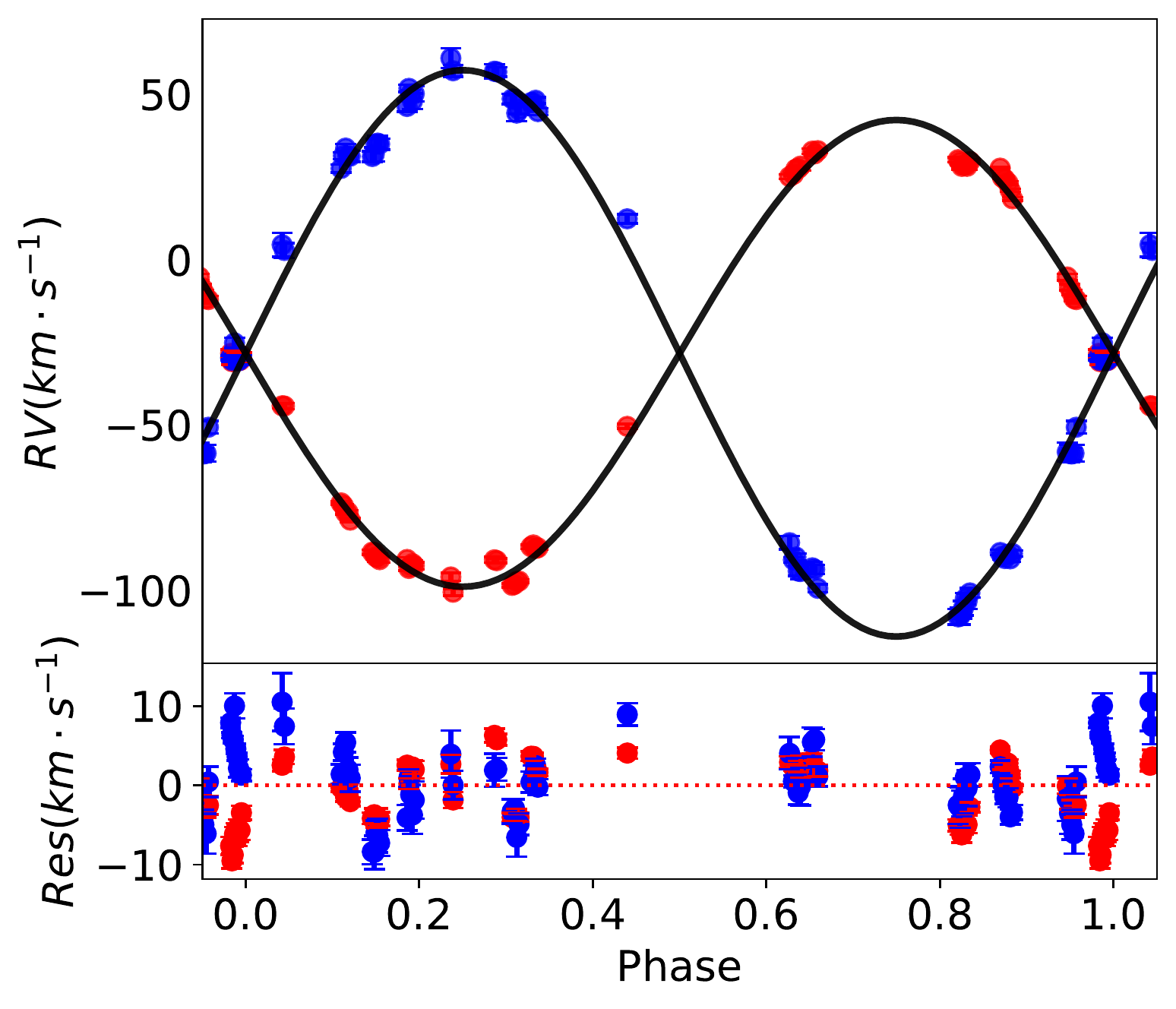}
  }
  \subfigure[]{
   \includegraphics[height=6cm, width=7cm]{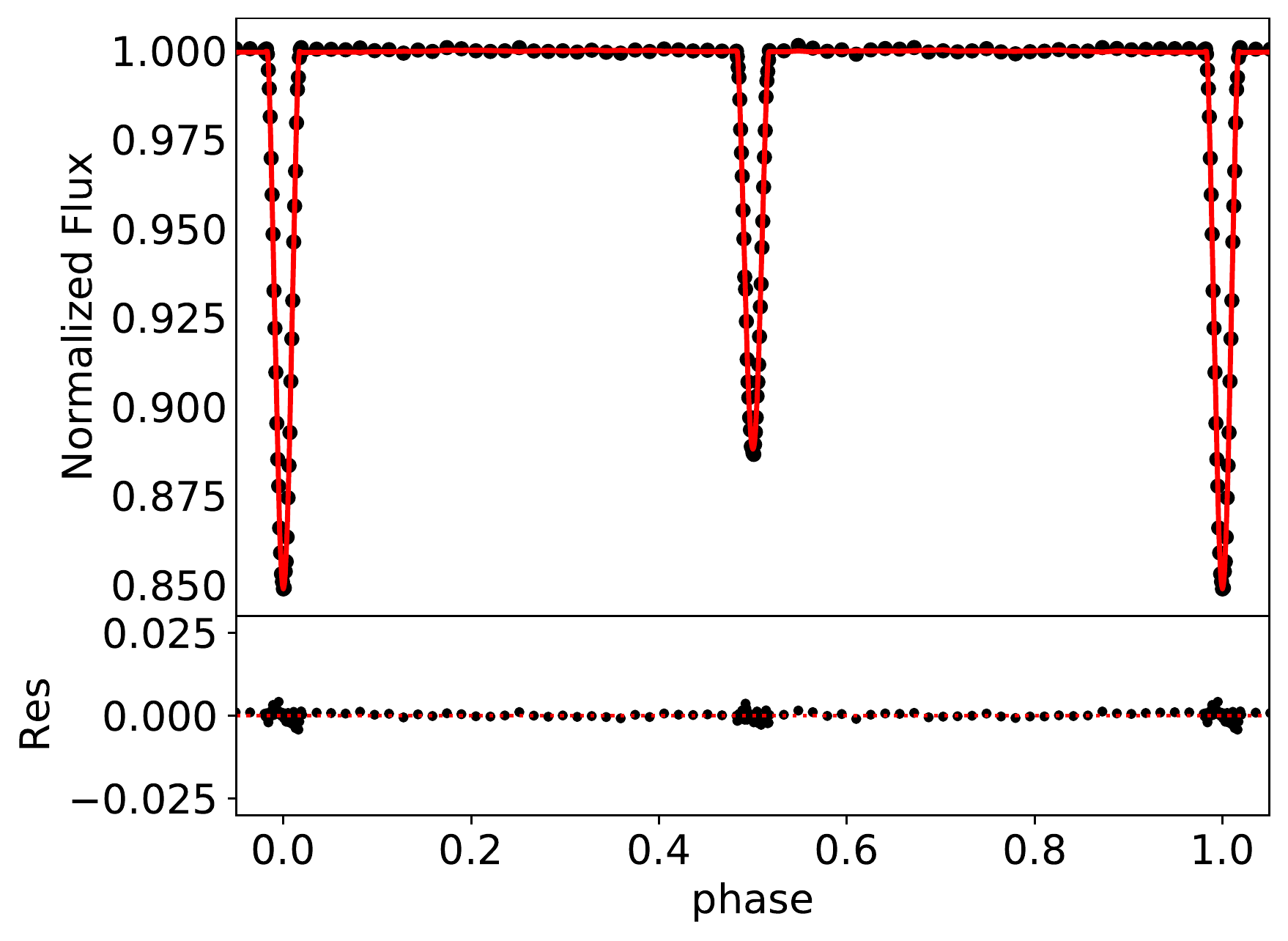}
  }
  \centering
  \caption{Panel (a): The radial velocity curves. In the top panel, the red and blue dots are the observed radial velocities of KIC 5359678 \citep{2021MNRAS.504.4302W}. The best-fit result derived by PHOEBE is shown with black solid lines. The bottom panel shows the residuals of radial velocities between the model and observations. Panel (b): The top panel shows the light curves for KIC 5359678 with black dots. The red solid line represent for the best-fit model derived by PHOEBE. The bottom panel shows the residuals between model and observations. }\label{fig:2}
\end{figure*}

\section{Method} \label{sec:method}
\subsection{Brief description}
For each star, the light curve with eclipses and the radial velocity curves of two companions covering the whole period are ready to derive the orbital solution.  
Light curves with eclipses are able to constrain the inclination and relative radii of both companions with respect to the semi-major axis ($R/a$). However, the stellar mass cannot be solely determined without double-line radial velocity curves. In this subsection we describe the general method to derive the stellar masses and radii of two companions in a binary system via the orbital solution.

The radial velocity of any of the companions in a spectroscopic binary system can be written as
\begin{small}
\begin{equation}\label{eq:RV}
    RV_{i} = \frac{2\pi a_i\sin i}{P(1-e^2)^{1/2}}[\cos(\theta+\omega)+e\cos \omega]+\gamma,
\end{equation}
\end{small}
where $P$ is the orbital period, $\theta$ is the angular position, $\omega$ is the longitude of the periastron, $e$ is the orbital eccentricity, $a_i$ is the semi-major distance from the $i$th companion to the barycenter, $i$ is the inclination of orbit, and $\gamma$ is the systematic velocity. The coefficient of the right-hand side is actually the semi-amplitude of the radial velocity curve and is usually denoted as
\begin{small}
\begin{equation}\label{eq:K}
\begin{aligned}
   K_{1}=&\frac{2\pi a_{1}\sin i}{P(1-e^2)^{1/2}},\\ K_{2}=&\frac{2\pi a_{2}\sin i}{P(1-e^2)^{1/2}}.
   \end{aligned}
\end{equation}
\end{small}

$K_i$ and $P$ are associated with the stellar masses via the mass function, which can be written as

\begin{small}
\begin{equation}\label{Equ:1}
\begin{aligned}
   f(M_{1})=&\frac{M_{1}^3\sin^3i}{(M_{1}+M_{2})^2}\\
   =&(1.0361\times10^{-7})(1-e^2)^{3/2}K_2^3P& M_\odot,\\
   f(M_{2})=&\frac{M_{2}^3\sin^3i}{(M_{1}+M_{2})^2}\\
   =&(1.0361\times10^{-7})(1-e^2)^{3/2}K_1^3P& M_\odot.
\end{aligned}
\end{equation}
\end{small}
Although the radial velocity curves can give $K_1$ and $K_2$, they cannot directly solve $M_1$ and $M_2$ without the help of light curves.

We then seek the orbit solution with the combination of the radial velocities and light curves using PHysics Of Eclipsing BinariEs (PHOEBE 2.2 \footnote{http://phoebe-project.org}; \citealt{2005ApJ...628..426P,2016ApJS..227...29P,2020ApJS..247...63J}), which is based on the WD code \citep{1971ApJ...166..605W}.

The atmospheric parameters of the F, G, and K type stars among the 56 stars are from the LAMOST pipeline \citep{2014IAUS..306..340W}, while those of the M dwarf and OB type stars are from \citet{2021ApJS..253...45L} and \citet{2021ApJS..257...54G}, respectively. For the secondary star, the \teff\,, \logg\, are resulted by the best orbit fitting from PHOEBE.

In this work, the effective temperature and \logg\, of the primary stars are estimated from the spectra observed near the secondary minimum eclipse, at which the secondary stars give minimum contribution in fluxes. As an instance, the secondary only contributes about 11\% of the primary fluxes near the secondary minimum eclipse when q=0.7 \citep{2018MNRAS.473.5043E}. These parameters are then dominated by primary stars and used as inputs in PHOEBE for orbit solution.

\citet{2018MNRAS.473.5043E} performed experiments with synthetic spectra to investigate the systematic biases of atmospheric parameters for unresolved main-sequence binaries on spectral fitting with single star models, they modeled spectra similar to those collected by the APOGEE, GALAH and LAMOST surveys. They found that, when an unresolved binary was considered as a single star, the typical errors of \teff\,, \logg\, and [Fe/H] estimates from the combined spectra are $<$200 K , $<$0.1 dex and $<$0.1 dex for LAMOST low-resolution spectra. These systematic errors are analog to the measurement error and hence would not significantly affect the final results of the mass and radius estimation.

A Markov chain Monte Carlo (MCMC \footnote{http://dan.iel.fm/emcee}; \citealt{2013PASP..125..306F}) sampling is then applied on PHOEBE to solve out all orbital parameters. The likelihood used in the MCMC is a joint chi-square value from the residuals of the fits with the observational light curve and radial velocity curves. The following priors are also added to restrict the MCMC sampling:
\begin{enumerate}
\item mass ratio ($q$): $0 < q < 2$
\item orbit inclination ($i^{\circ}$): $0 < i < 90^{\circ}$
\item eccentricity ($e$): $0 < e < 1$
\item argument of periastron ($\omega^{\circ}$): $0 < \omega < 360^{\circ}$
\item semi-major of orbit ($a$): $0 < a/R_\odot < 100$
\item system velocity ($V_{\gamma}:$): $-200 < V_{\gamma}/({\rm km\,s^{-1}}) < 200$
\item effective temperature of secondary ($T_{2}$): $3000 < T_{2}/{\rm K} < T_{2}+3\sigma_{T_{1}}$
\item equivalent radius of primary ($R_{1}$): $0 < R_{1}/R_{\odot} < 10$
\item equivalent radius of secondary ($R_{2}$): $0 < R_{2}/R_{\odot} < 10$
\end{enumerate}

\begin{table*}[t]
\caption{Comparison of the orbital parameters of KIC 5359678 between this work and \citet{2021MNRAS.504.4302W}.}\label{tab:table1}
\begin{tabular}{ccc|ccc}
\hline
\hline
  \multicolumn{1}{c}{Parameters} &
  \multicolumn{1}{c}{\citep{2021MNRAS.504.4302W}} &
  \multicolumn{1}{c|}{This work} &
  \multicolumn{1}{c}{Parameters} &
  \multicolumn{1}{c}{\citep{2021MNRAS.504.4302W}} &
  \multicolumn{1}{c}{This work}\\
\hline
\hline
$e$ & 0.00032 $\pm$ 0.00006 & 0.0024 $\pm$ 0.0057 &$T_{2} (K)$ & 5980 $\pm$ 22 & 6040.84 $\pm$ 40.75\\
\hline
$\omega$ $^{\rm a}$ $(\circ)$ & -89.55 $\pm$ 1.05 & 88.00 $\pm$ 6.03 &$M_{1} (M_{\odot})$ & 1.320 $\pm$ 0.060  & 1.361 $\pm$ 0.008\\
\hline
$i $ $(\circ)$ & 85.56 $\pm$ 0.10 & 85.56 $\pm$ 0.37 &$M_{2} (M_{\odot})$ & 1.12 $\pm$  & 1.121 $\pm$ 0.008\\
\hline
$q $ & 0.851 $\pm$ 0.10 & 0.825 $\pm$ 0.011 &$R_{1} (R_{\odot})$ & 1.52 $\pm$ 0.04  & 1.490 $\pm$ 0.080\\
\hline
$V_{\gamma}$ (\kms) & -29.26 $\pm$ 0.19 & -28.29 $\pm$ 0.34 &$R_{2} (R_{\odot})$ & 1.05 $\pm$0.05 & 1.075 $\pm$ 0.130\\
\hline
$a (R_{\odot})$ & 19.19    & 19.28 $\pm$ 0.13 &$log\emph{g}$ & - & 4.200 $\pm$ 0.079 \\
\hline
$T_{1} (K)$ & 6500 $\pm$ 50  & 6501.04 $\pm$ 47.94 &$[M/H]$ & - & -0.104 $\pm$ 0.046\\
\hline
\end{tabular}
\tabnote{$^{\rm a}$\citet{2021MNRAS.504.4302W} derived $\omega$ by ecos$\omega$=$\pi$/2[($\phi_{2}-\phi_{1}$)-0.5], and esin$\omega$=(w$_{2}$-w$_{1}$)/(w$_{2}$-w$_{1}$). $\phi_{2}$ is the phase of the secondary eclipses, $\phi_{1}$=0, w$_{1}$ and w$_{2}$ are widths of primary and secondary eclipse in phase, respectively\citep{2015PASA...32...23K}.} 
\end{table*}

\begin{table*}[t]
\scriptsize
\caption{List of 10 of the samples compiled from previous studies with accurate masses, radii, and atmospheric parameters (\teff, \logg, [M/H])}\label{tab:table2}
\centering
\begin{tabular}{cccccccc}
\hline
\hline
  \multicolumn{1}{c}{Name} &
  \multicolumn{1}{c}{\teff\,[K]} &
  \multicolumn{1}{c}{\logg\,[dex]} &
  \multicolumn{1}{c}{[M/H]\,[dex]} &
  \multicolumn{1}{c}{$\log{L/L_\odot}$} &
  \multicolumn{1}{c}{M\,[$M_{\odot}$]} &
  \multicolumn{1}{c}{R\,[$R_{\odot}$]} &
  \multicolumn{1}{c}{\textbf{Ref.}} \\
\hline
\hline
47 Tuc E32 & 6025.60 $\pm$ 18.08 & 4.227$\pm$ 0.002 & -0.71 $\pm$ 0.10 & 0.221 $\pm$ 0.012 & 0.862 $\pm$ 0.002 & 1.183 $\pm$ 0.001 & \citep{2020MNRAS.492.4254T}\\
& 5956.62 $\pm$ 17.87 & 4.352 $\pm$ 0.003 & -0.71 $\pm$ 0.10 & 0.059 $\pm$ 0.012 & 0.827 $\pm$ 0.002 & 1.004 $\pm$ 0.002 &\\
\hline

47 Tuc V69  &  5956.62 $\pm$ 17.86 &  4.143 $\pm$ 0.003  & -0.71 $\pm$ 0.1  & 0.293 $\pm$ 0.012  & 0.876 $\pm$ 0.002  & 1.315 $\pm$ 0.002 &\citep{2020MNRAS.492.4254T}\\
& 5984.12 $\pm$ 17.95 & 4.242 $\pm$ 0.003 & -0.71 $\pm$ 0.10 & 0.193 $\pm$ 0.013 & 0.859 $\pm$ 0.003 & 1.162 $\pm$ 0.003 &\citep{2017MNRAS.468..645B} \\
\hline

AD Boo & 6575.0 $\pm$ 120.0 & 4.173$\pm$ 0.008 & 0.10$\pm$ 0.15 & 0.642 $\pm$ 0.075 & 1.414 $\pm$ 0.008 & 1.613 $\pm$ 0.014 & \citep{2008AA...487.1095C} \\
& 6145.0 $\pm$ 120.0 & 4.350 $\pm$ 0.007 & 0.10$\pm$ 0.15 & 0.279 $\pm$ 0.080 & 1.209 $\pm$ 0.006 & 1.216 $\pm$ 0.010 &\\
\hline

AH Cep & 30690.22 $\pm$ 245.52 & 4.019 $\pm$ 0.012 & 0.0 $\pm$ 0.0 & 4.530 $\pm$ 0.034 & 16.140 $\pm$ 0.113 & 6.510 $\pm$ 0.044&\citep{1991AJ....101..600P}\\
& 28773.98 $\pm$ 230.19 & 4.073 $\pm$ 0.018 & 0.0$\pm$0.0 & 4.294 $\pm$0.036 & 13.690 $\pm$0.092 & 5.640 $\pm$ 0.048&\citep{2018MNRAS.481.3129P}\\
\hline

AI Phe &5010.0 $\pm$ 120.0 & 3.595$\pm$ 0.014 & -0.14 $\pm$ 0.10& 0.689 $\pm$ 0.101&1.234 $\pm$ 0.005 & 2.932 $\pm$ 0.048&\citep{1988AA...196..128A}\\
 & 6310.0 $\pm$ 150.0 & 3.996 $\pm$ 0.011 &-0.14 $\pm$ 0.10 & 0.674 $\pm$ 0.098&1.193 $\pm$ 0.004 & 1.818 $\pm$ 0.024 &\\
\hline

AL Ari & 6367.95 $\pm$ 25.47 & 4.229 $\pm$ 0.005 & -0.42 $\pm$ 0.08 & 0.446 $\pm$ 0.017 & 1.164 $\pm$ 0.001 & 1.372 $\pm$ 0.004&\citep{2021AA...649A.109G}\\
& 5559.04$\pm$ 27.80 & 4.458 $\pm$ 0.008 &-0.42 $\pm$ 0.08 & -0.152 $\pm$ 0.021 & 0.911 $\pm$ 0.000 & 0.905 $\pm$ 0.003&\\
\hline

AL Dor & 6053.41 $\pm$ 30.27 & 4.404 $\pm$ 0.002 & -0.10 $\pm$ 0.04 & 0.159 $\pm$ 0.020 & 1.102$\pm$ 0.000 & 1.092 $\pm$ 0.001&\citep{2019AA...632A..31G}\\
 & 6053.41$\pm$ 30.27 & 4.399 $\pm$ 0.002 &  -0.10 $\pm$ 0.04 & 0.164 $\pm$ 0.020 & 1.103 $\pm$ 0.000 & 1.098 $\pm$ 0.001&\citep{2021AA...649A.109G}\\
\hline

ASAS J045021+2300.4 & 5662.39 $\pm$28.31 & 4.360 $\pm$ 0.018 & -0.26 $\pm$ 0.26 & 0.016 $\pm$ 0.027 & 0.934 $\pm$ 0.007 & 1.058$\pm$ 0.010&\citep{2021MNRAS.508.5687H}\\
 & 3589.22 $\pm$ 43.07 & 4.829 $\pm$ 0.019 & -0.26 $\pm$ 0.26& -1.604 $\pm$ 0.052 & 0.409$\pm$ 0.002 & 0.408 $\pm$ 0.004&\\
\hline

ASAS J051753-5406.0 & 5984.12 $\pm$ 89.76 & 3.982 $\pm$ 0.006 & -0.1 $\pm$ 0.13 & 0.636 $\pm$ 0.061 & 1.311$\pm$ 0.015 & 1.935 $\pm$0.009&\citep{2022MNRAS.tmp.2880M}\\
 & 5847.90 $\pm$ 81.87 & 4.332 $\pm$ 0.008 & -0.1 $\pm$ 0.13 & 0.167 $\pm$ 0.057 & 1.093 $\pm$ 0.013 & 1.181 $\pm$ 0.006&\\
\hline

ASAS J052821+0338.5 & 5105.05 $\pm$ 45.95 & 4.05 $\pm$ 0.01 & -0.15 $\pm$ 0.14 & 0.312 $\pm$0.036 & 1.375 $\pm$ 0.005 & 1.830$\pm$ 0.004&\citep{2008AA...481..747S}\\
& 4709.77 $\pm$ 42.39 & 4.08 $\pm$ 0.01 & -0.15 $\pm$ 0.14 & 0.123 $\pm$ 0.036 & 1.329 $\pm$ 0.003 & 1.730$\pm$ 0.004&\\

\hline
\end{tabular}
\end{table*}

We initialized the chains of MCMC using random values drawn from uniform distributions with the restricting ranges defined above. The surface gravity and reflection coefficient are fixed in PHOEBE unless the \teff\ of the primary star is larger than 8000\,$K$. For each star, we run the model with 80 walkers, each of which takes 2000 steps. Finally, we calculate the peak value of the probability distribution and the standard deviations of the random drawn points as the best-fit parameters and their uncertainties.

\subsection{Method verification}
As a verification of the method, we solve out the orbit for KIC 5359678 and compare with the results with \citep{2021MNRAS.504.4302W}.
Figure \ref{fig:1} shows our MCMC result for KIC 5359678. In the corner plot, each histogram shows the probability distributions of $q$, $i$, $\sqrt{e}cos\omega$, $\sqrt{e}sin\omega$,$a$,  $V_{\gamma}$, $T_{2}$, $R_{1}$ and $R_{2}$. 

Figure \ref{fig:2} (a) shows the observed and the best-fit radial velocity curve. The residuals of the velocity indicate that the uncertainty of the best-fit radial velocity curve is around 4.24\,\kms. The top of panel (b) shows the best-fit light curve model by PHOEBE (red line) and the corresponding light observation (black dots), and the bottom of the panel shows the residuals of the light curve between the model fit and observations. It illustrates that the best-fit radial velocity curves model and light curve well match with the observed data.  

The masses and radii of the primary and secondary of KIC5359678 that we obtain from PHOEBE with MCMC are $M_{1} = 1.361 \pm 0.008M_{\odot}$, $R_{1} = 1.49 \pm 0.08R_{\odot}$ $M_{2} = 1.121 \pm 0.008M_{\odot}$, $R_{2} = 1.075 \pm 0.130 R_{\odot}$. 
All best-fit parameters of the MCMC and those obtained by \citep{2021MNRAS.504.4302W} are listed and compared in Table \ref{tab:table1}.
It illustrates that the orbital parameters obtained with PHOEBE+MCMC are consistent with those by \citet{2021MNRAS.504.4302W} within the uncertainties.

\begin{figure}[t]
  \centering
   \includegraphics[height=6cm, width=7cm]{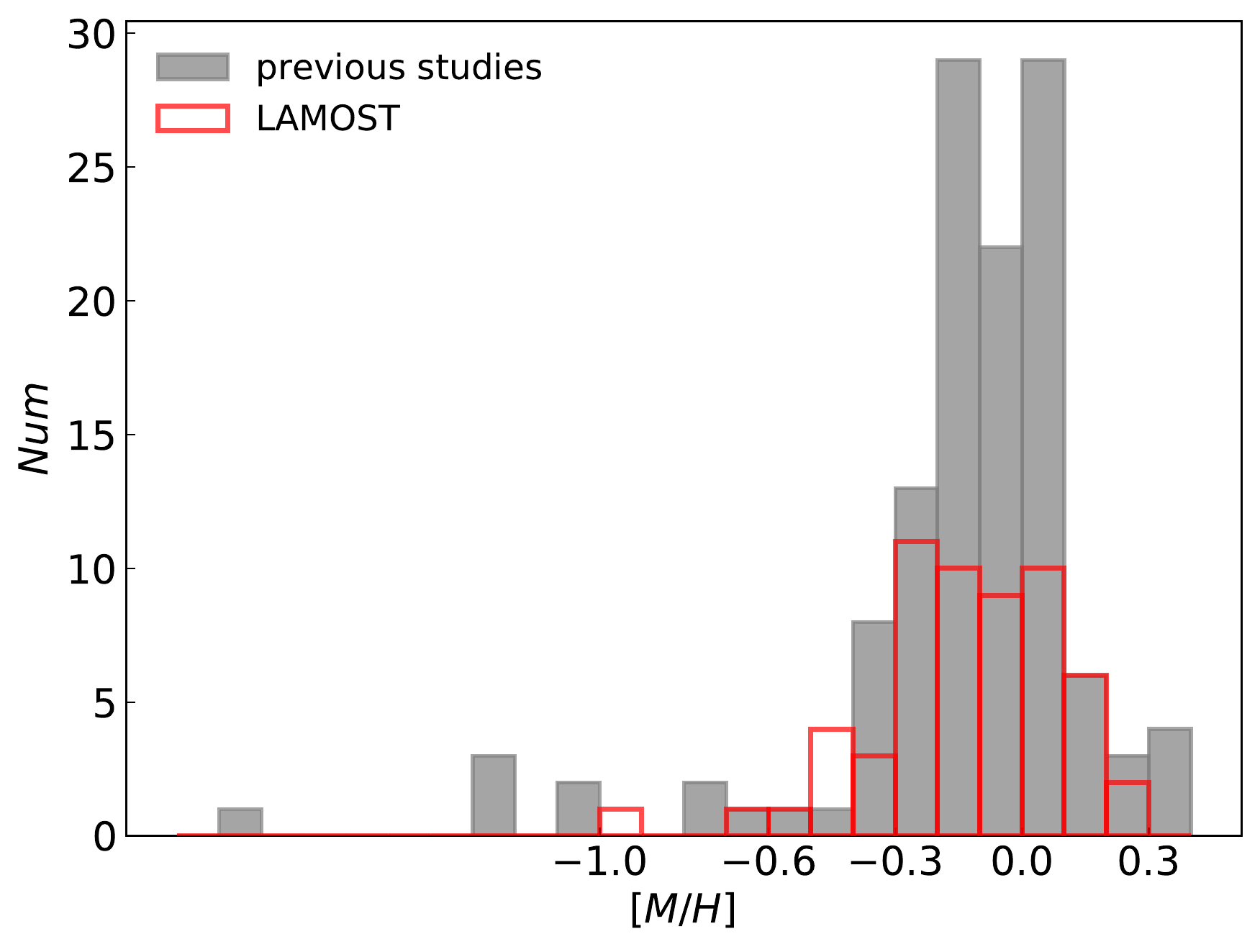}
  \centering
  \caption{The\mh\ distribution of sample stars compiled in this work.}\label{fig:fig3}
\end{figure}

\begin{table*}[t]
\scriptsize
\caption{Summary of 10 of binary samples obtained from LAMOST MRS with the masses, radii, and atmospheric parameters (\teff, $\log{g}$, [M/H]).}\label{tab:table3}
\centering
\begin{tabular}{ccccccccc}
\hline
\hline
  \multicolumn{1}{c}{No.} &
  \multicolumn{1}{c}{\textbf{gaia\_source\_id}} &
  \multicolumn{1}{c}{\teff\,[K]} &
  \multicolumn{1}{c}{$\log{g}$\,[dex]} &
  \multicolumn{1}{c}{$\log{L/L_\odot}$} &
  \multicolumn{1}{c}{$M$\,[$M_{\odot}$]} &
  \multicolumn{1}{c}{$R$\,[$R_{\odot}$]} &
  \multicolumn{1}{c}{[M/H]\,[dex]} &
  \multicolumn{1}{c}{$v\sin{i}$[\kms]} \\
\hline
\hline
$1^{\footnote{\citep{2021MNRAS.504.4302W}}}$&2101510803402761344&6500.00$\pm$50.00&4.200$\pm$0.079&0.570$\pm$0.061&1.320$\pm$0.060&1.520$\pm$0.040&-0.104$\pm$0.046&79.310\\
 &&5980.00$\pm$22.00&4.544$\pm$0.070&0.104$\pm$0.096&1.120$\pm$0.000&1.050$\pm$0.050&-0.104$\pm$0.046&-\\
$2^{\footnote{\citep{Pan_2020}}}$&2126356983051726720&6144.00$\pm$100.00&4.220$\pm$0.010&0.431$\pm$0.066&1.290$\pm$0.020&1.450$\pm$0.010&-0.019$\pm$0.035&288.310\\
 &&5966.00$\pm$97.00&4.330$\pm$0.020&0.216$\pm$0.067&1.110$\pm$0.050&1.200$\pm$0.010&-0.019$\pm$0.035&-\\
3&2101192082470870912&6012.74$\pm$23.74&4.105$\pm$0.022&0.413$\pm$0.144&1.208$\pm$0.003&1.482$\pm$0.106&0.142$\pm$0.014&89.190\\
 &&5767.75$\pm$21.55&3.985$\pm$0.051&0.559$\pm$0.088&1.281$\pm$0.003&1.905$\pm$0.083&0.142$\pm$0.014&-\\
4&3811041442290209024&5940.85$\pm$32.41&4.157$\pm$0.053&0.237$\pm$0.164&1.065$\pm$0.031&1.240$\pm$0.101&-0.303$\pm$0.031&54.630\\
 &&4482.10$\pm$131.76&3.547$\pm$0.080&0.463$\pm$0.222&1.027$\pm$0.031&2.824$\pm$0.266&-0.303$\pm$0.031&-\\
5&3423803170795356800&16074.66$\pm$100.00&3.656$\pm$0.300&2.838$\pm$0.059&3.452$\pm$0.044&3.384$\pm$0.091&-0.350$\pm$0.100&-\\
 &&16186.85$\pm$200.69&4.019$\pm$0.100&2.724$\pm$0.143&3.272$\pm$0.044&2.927$\pm$0.197&-0.350$\pm$0.100&-\\
6&3378228658638299136&9824.11$\pm$1000.00&3.659$\pm$0.100&1.676$\pm$0.423&2.137$\pm$0.060&2.375$\pm$0.136&-0.482$\pm$0.100&-\\
 &&9423.69$\pm$249.40&4.024$\pm$0.121&1.544$\pm$0.218&1.901$\pm$0.060&2.219$\pm$0.211&-0.482$\pm$0.100&-\\
7&604716006409635456&5529.74$\pm$45.16&4.210$\pm$0.100&0.238$\pm$0.320&1.594$\pm$0.013&1.432$\pm$0.228&0.030$\pm$0.130&66.400\\
 &&6598.67$\pm$519.58&4.000$\pm$0.238&0.826$\pm$0.506&1.432$\pm$0.013&1.980$\pm$0.392&0.030$\pm$0.130&-\\
8&598940203109496064&4697.68$\pm$43.23&4.130$\pm$0.200&-0.717$\pm$0.254&0.746$\pm$0.011&0.661$\pm$0.083&-0.224$\pm$0.170&-\\
 &&4658.78$\pm$123.60&4.434$\pm$0.260&-0.490$\pm$0.410&0.756$\pm$0.011&0.873$\pm$0.173&-0.224$\pm$0.170&-\\
9&609180023619173632&5971.97$\pm$10.79&3.800$\pm$0.040&0.696$\pm$0.101&1.148$\pm$0.009&2.081$\pm$0.105&-0.234$\pm$0.010&126.360\\
 &&5665.94$\pm$84.29&3.836$\pm$0.078&0.633$\pm$0.161&1.157$\pm$0.009&2.150$\pm$0.161&-0.234$\pm$0.010&-\\
10&658309364243893504&5840.19$\pm$128.78&3.972$\pm$0.107&0.688$\pm$0.311&0.825$\pm$0.030&2.156$\pm$0.322&0.034$\pm$0.067&83.920\\
 &&5343.71$\pm$240.28&3.770$\pm$0.153&0.468$\pm$0.378&0.859$\pm$0.030&1.998$\pm$0.332&0.034$\pm$0.067&-\\

  \hline
\end{tabular}
\end{table*}

\section{Results} \label{sec:result}
\begin{figure*}[!t]
  \centering
    \subfigure{
   \includegraphics[scale=0.34]{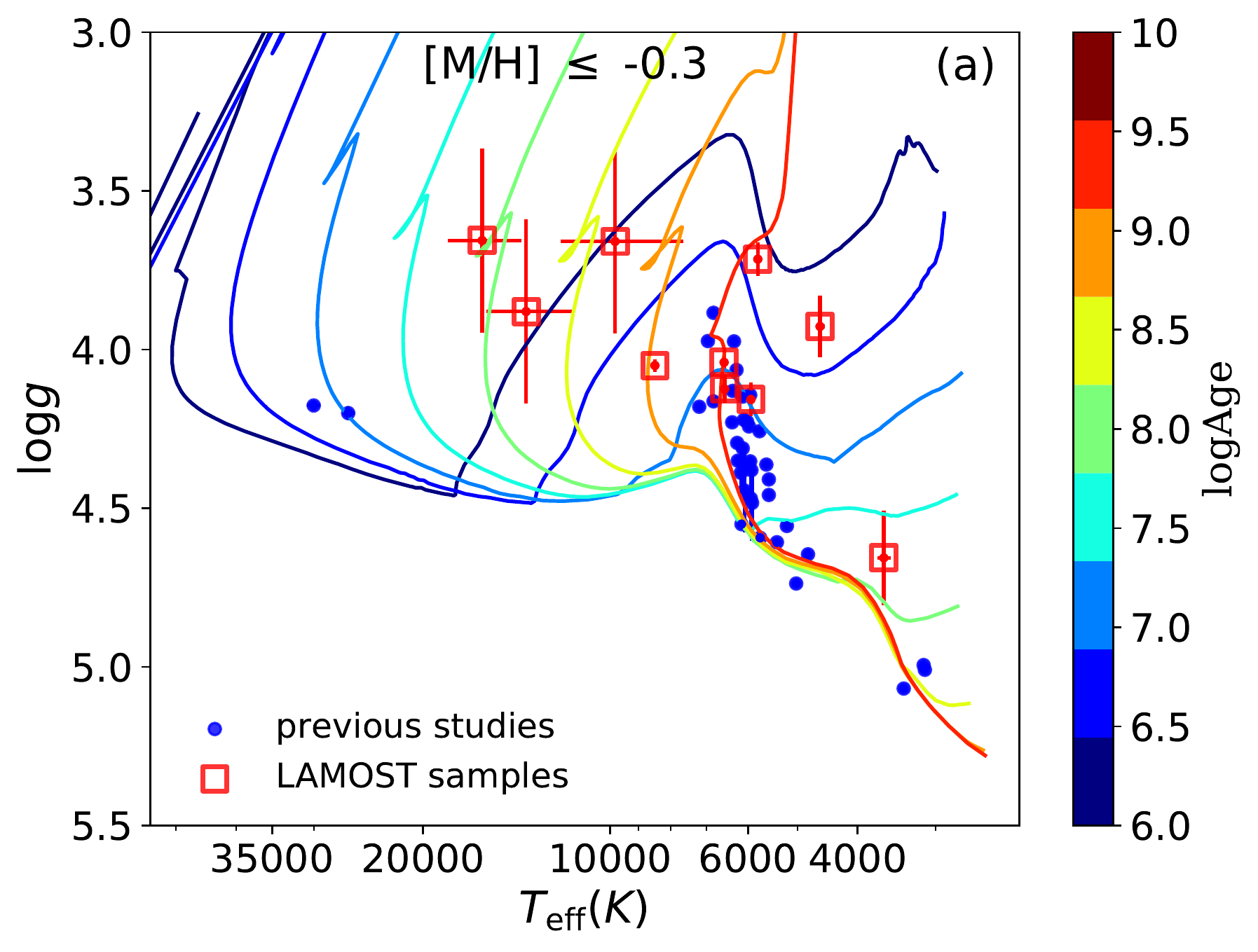}
  }
  \subfigure{
   \includegraphics[scale=0.34]{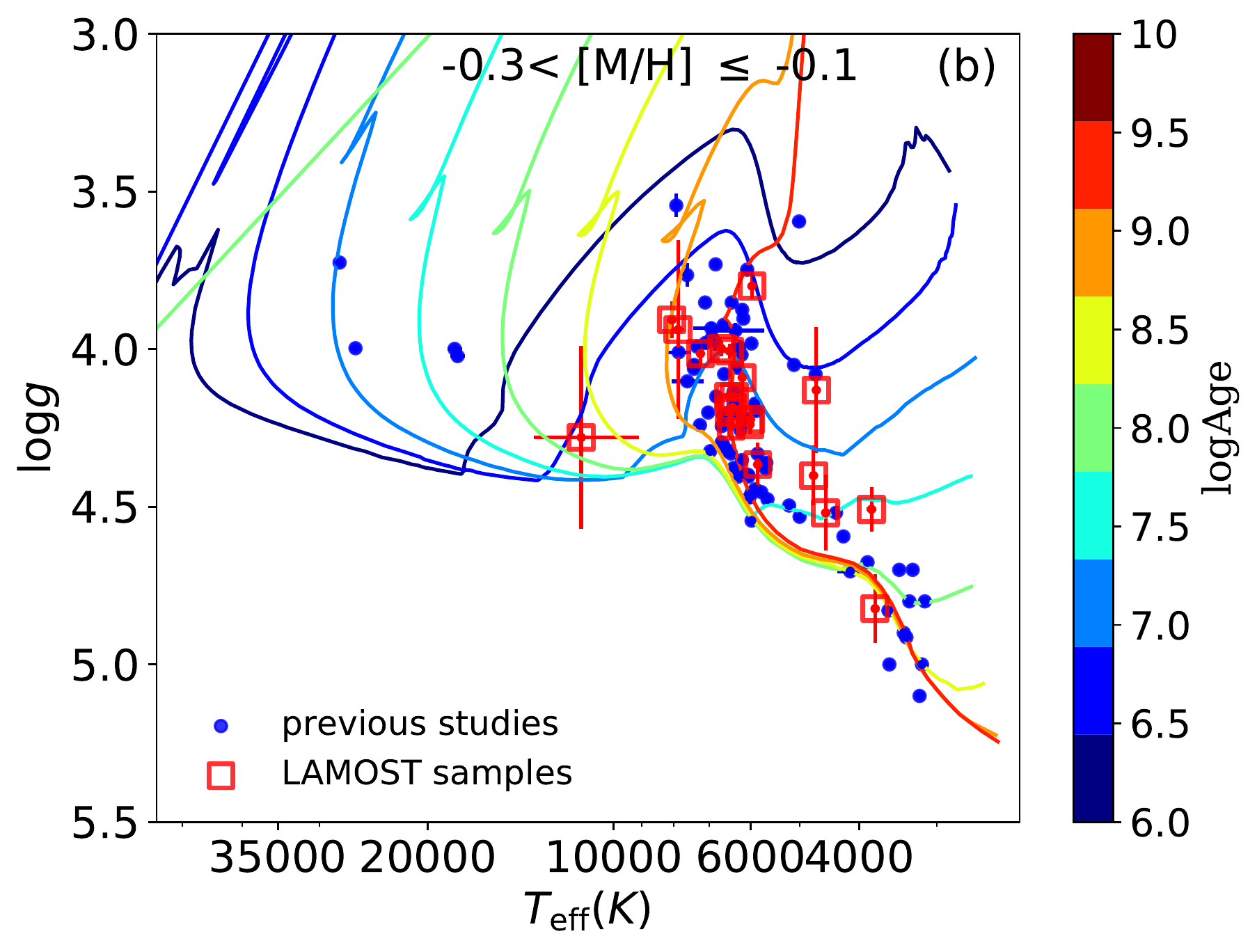}
  }
    \subfigure{
   \includegraphics[scale=0.34]{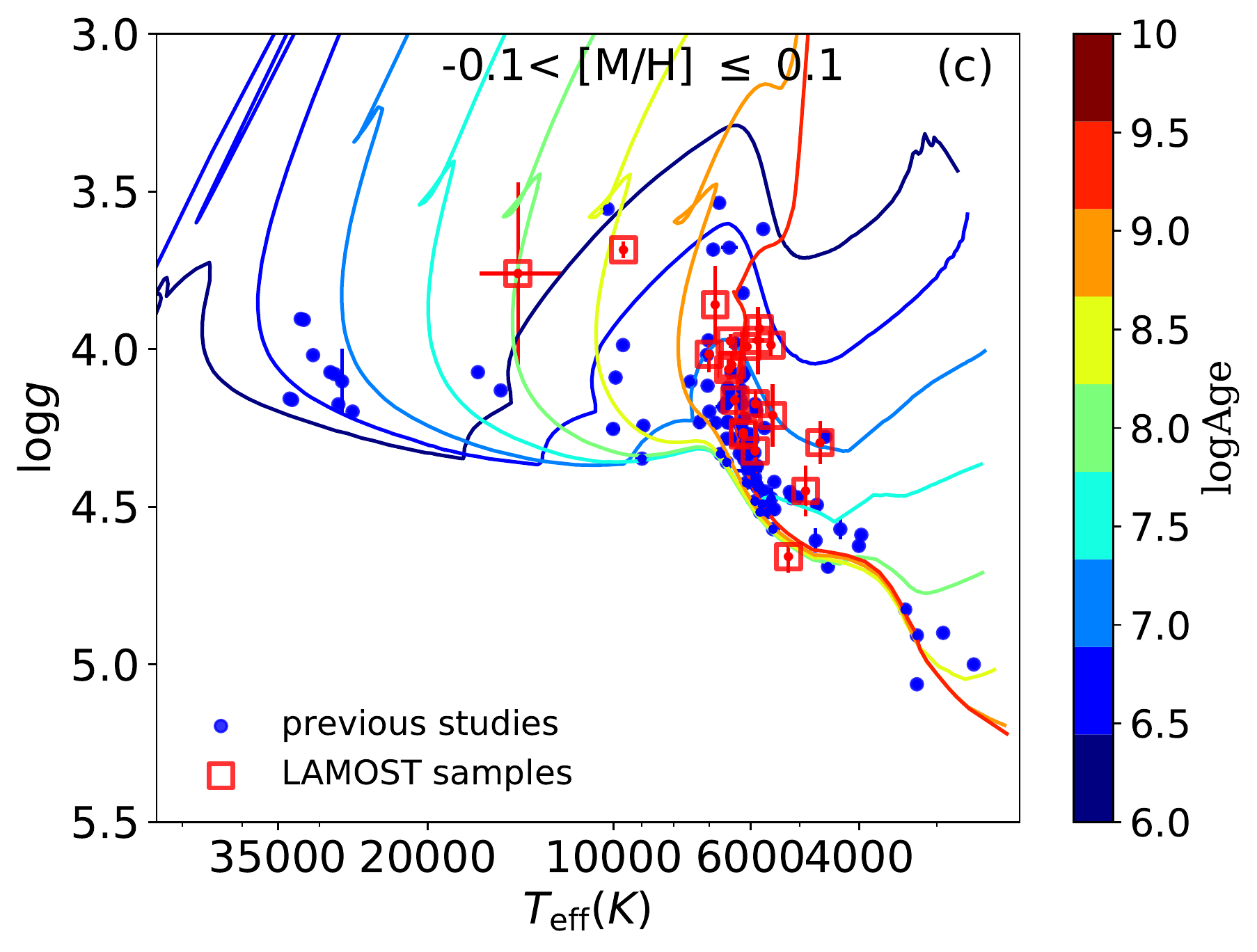}
  }
  \subfigure{
   \includegraphics[scale=0.34]{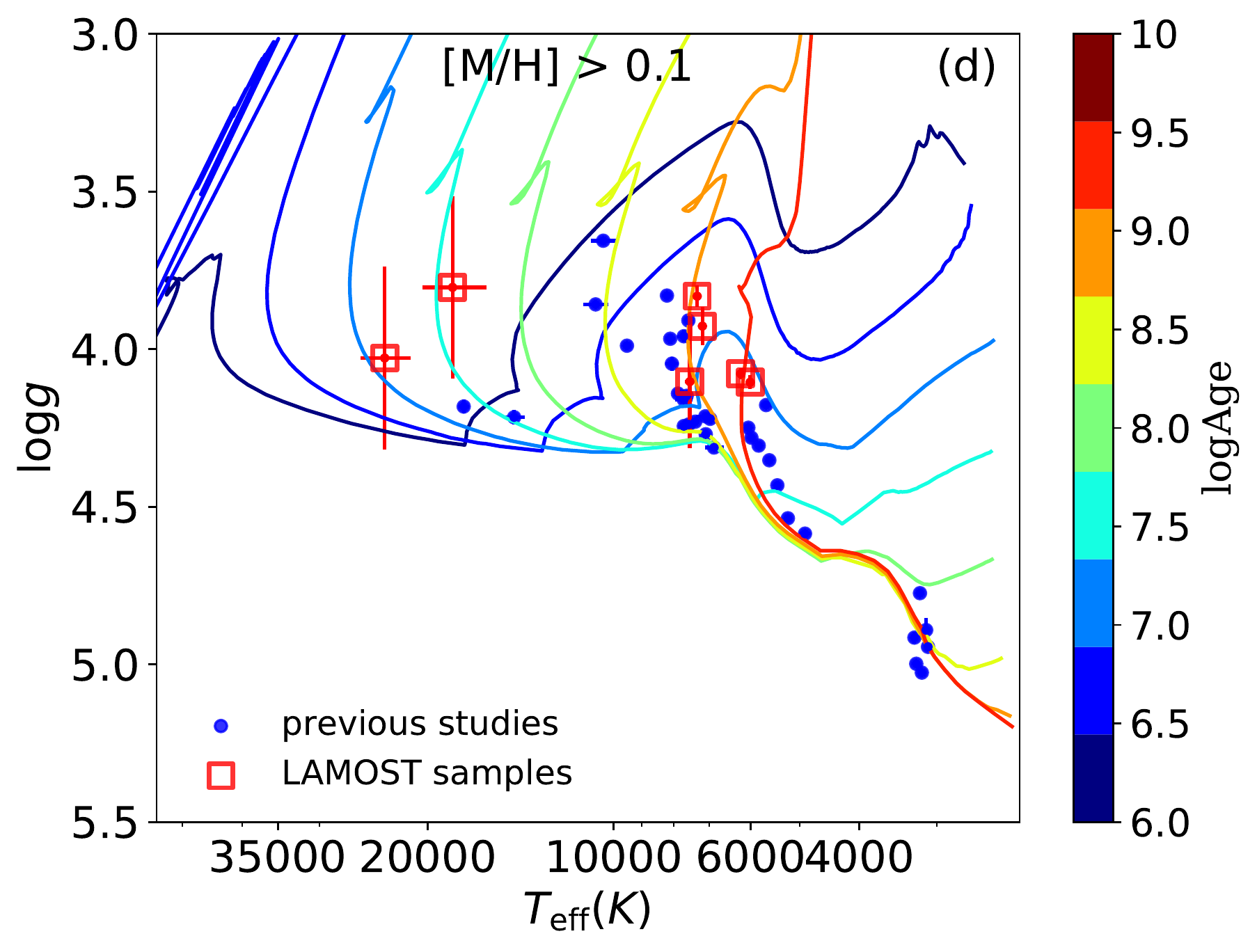}
  }
  \centering
  \caption{The (\teff-\logg) diagram of samples. From left to right and top to bottom, the panels display the samples with [M/H]$\leq-0.3$, $-0.3<$[M/H]$\leq-0.1$, $-0.1<$[M/H]$\leq+0.1$, and [M/H]$>+0.1$, respectively. The blue dots are the samples compiled from the literature and the red rectangles are the samples from LAMOST MRS. The lines are the isochrones given by PARSEC models \citep{2012MNRAS.427..127B} with color-coded ages ranging from 10\,Myr to 10\,Gyr.}\label{fig:fig4}
\end{figure*}
We compile 128 detached binary systems with atmospheric parameters and accurate masses and radii. The 10 of 128 detached eclipsing binaries are presented in table \ref{tab:table2}, the whole catalog can be found in China-VO: \dataset[doi:10.12149/101147]{https://doi.org/10.12149/101147}.

In table \ref{tab:table3}, the masses, radii, and atmospheric parameters of 58 double-line eclipsing binaries selected from LAMOST MRS are displayed. It contains 56 new orbital solutions of binaries. In table \ref{tab:table3}, we also provide the parameters of secondary stars obtained from orbital solution with PHOEBE. Although the internal errors of the parameters of the secondary stars are similar to those of the primary, \teff\ and \logg\ may be have larger uncertainty in the measurement. 

In this work, a total of 128 binaries with dynamic masses, radii, \teff, \logg, and [M/H] are compiled, including 128 binaries composed of two main-sequence stellar companions from literature data and 56 new orbited solutions of binaries from LAMOST DR8 MRS. Figure \ref{fig:fig3} shows the [M/H] distribution of these stars. The range of [M/H] is between -1.86 dex to 0.61 dex. The filled histogram is the [M/H] distribution of samples from literature, while the red hollow histogram indicates the distribution of the samples from LAMOST DR8 MRS. It is seen that, while most of the LAMOST samples are concentrated around the solar metallicity, similar to the previous samples, a few of them locating at lower metallicity help to densifying the distribution of metallicity.

Figure \ref{fig:fig4} shows the \teff-\logg\ relation of the stars in different \mh\ bins. Most of the samples are in the main-sequence stage. At the temperature around 6000 K, some older stars just about to move away from the main-sequence and show the turn-off point. The red rectangles show that the new LAMOST samples well fill up with the metal-poor regime (see the top-left panel). Meanwhile, because the hot stars are more sensitive to the age, the new LAMOST samples also provide more sampling point at different ages in the regime of high \teff.

\subsection{Comparison with PARSEC}

\begin{figure*}[!t]
  \centering
  \subfigure{
   \includegraphics[scale=0.38]{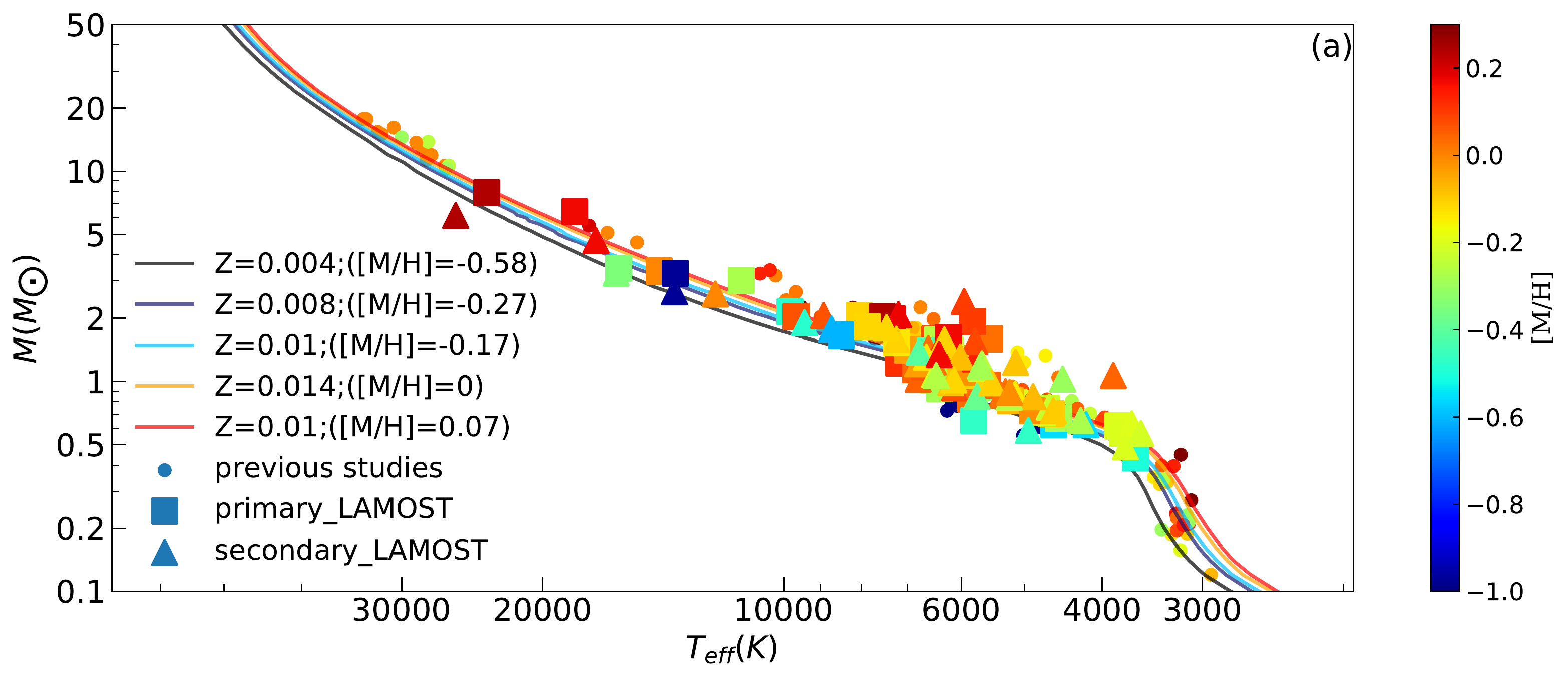}
  }
  \subfigure{
   \includegraphics[scale=0.34]{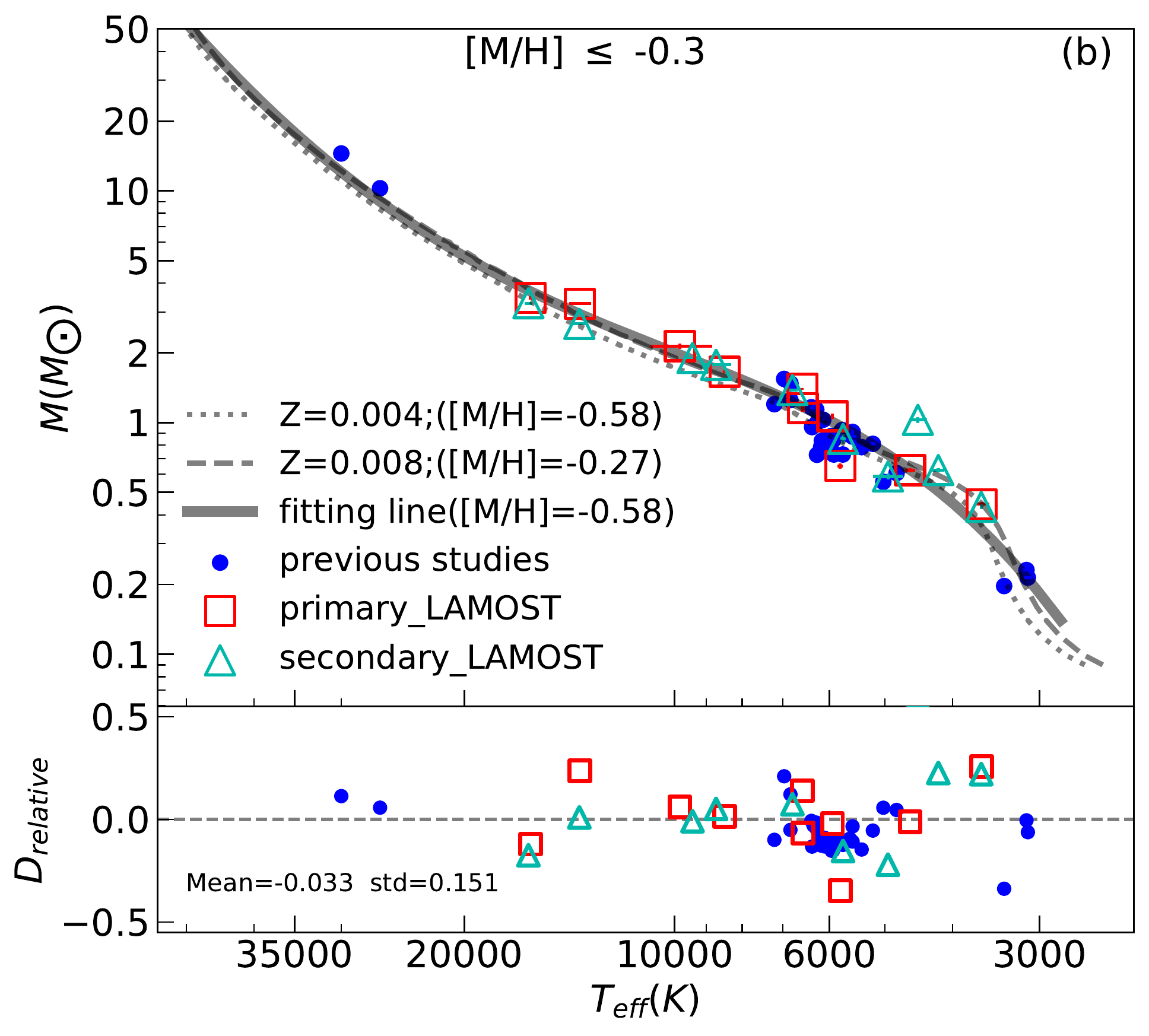}
  }
  \subfigure{
   \includegraphics[scale=0.34]{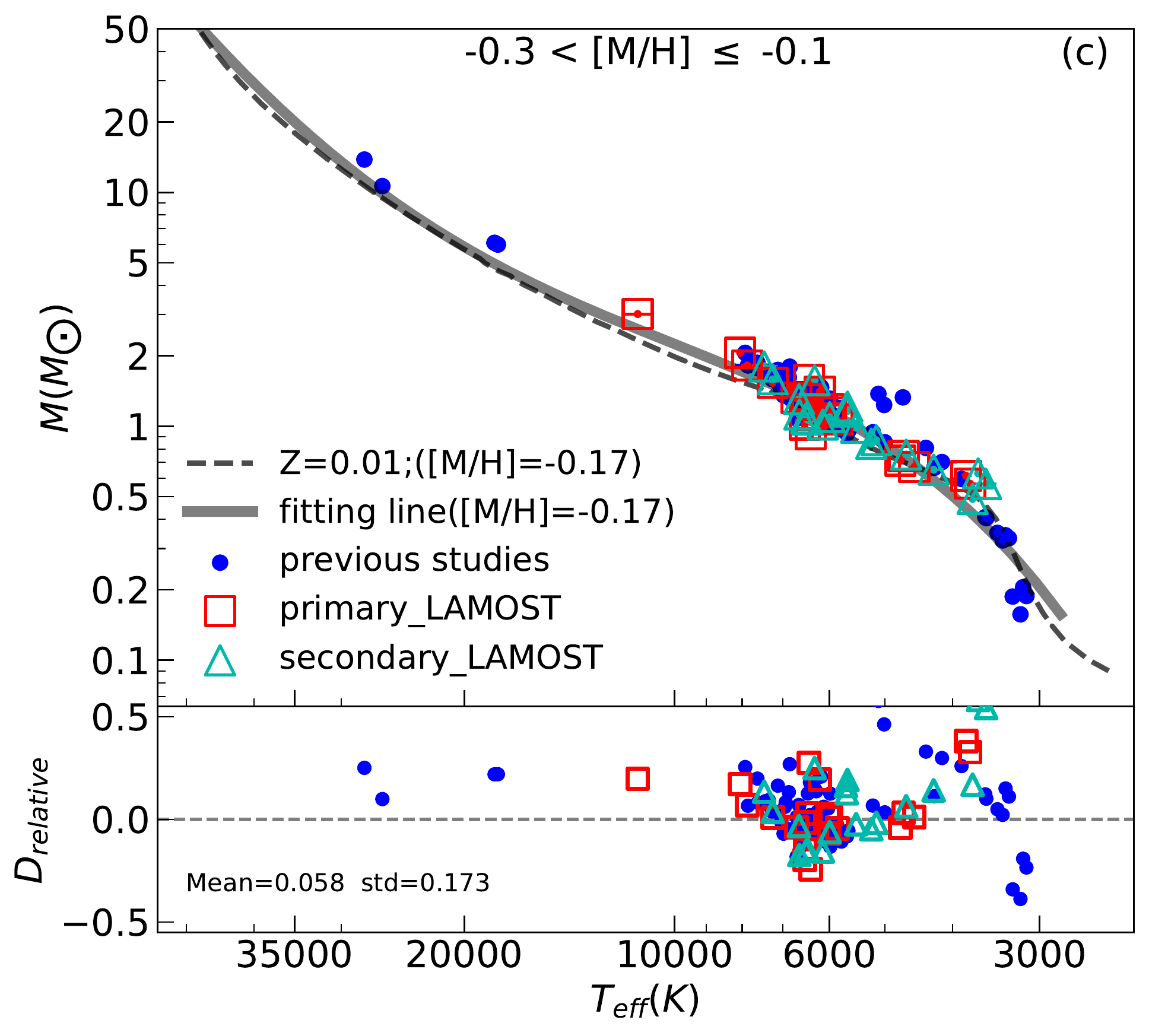}
  }
  \subfigure{
   \includegraphics[scale=0.34]{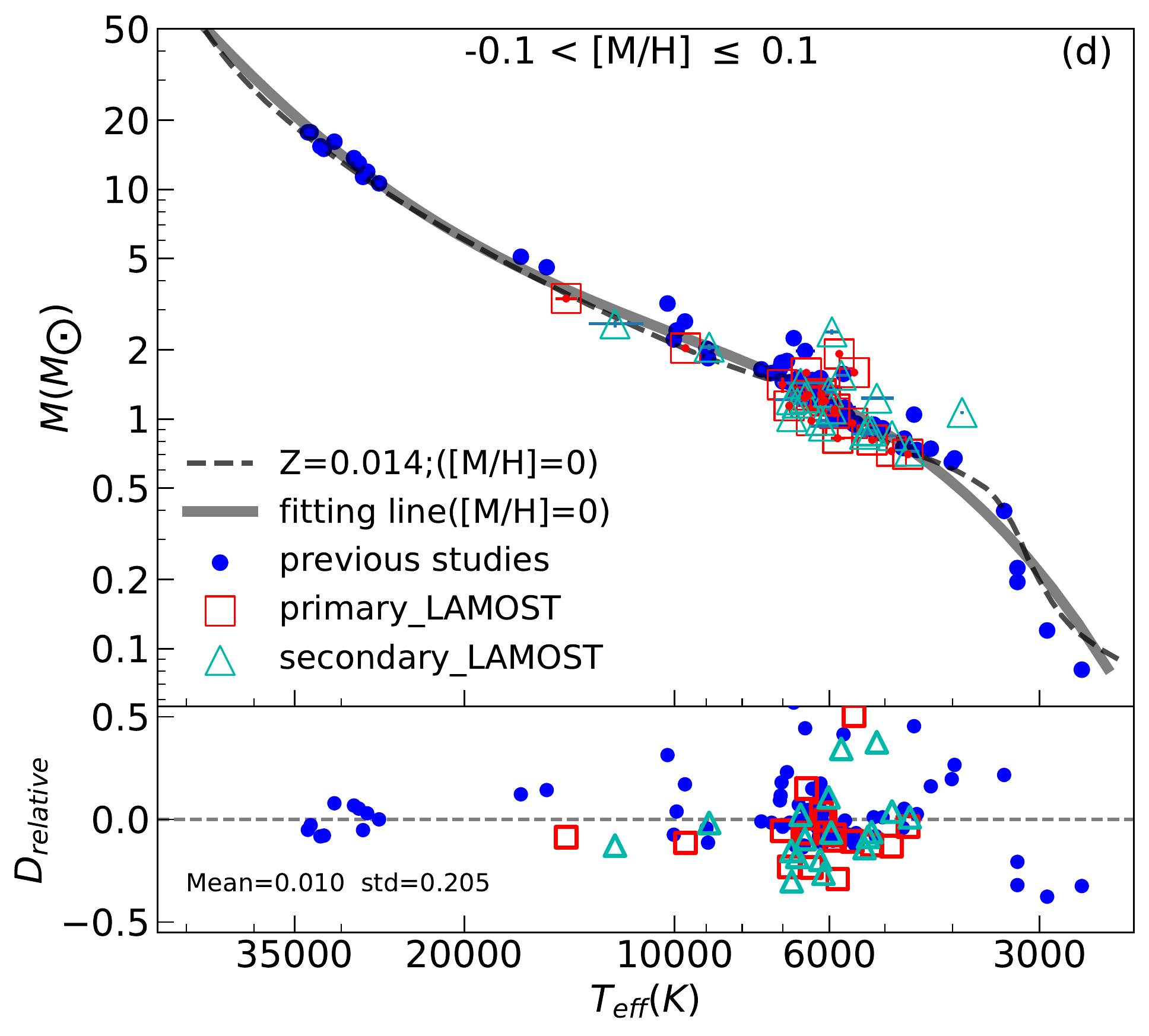}
  }
  \subfigure{
   \includegraphics[scale=0.34]{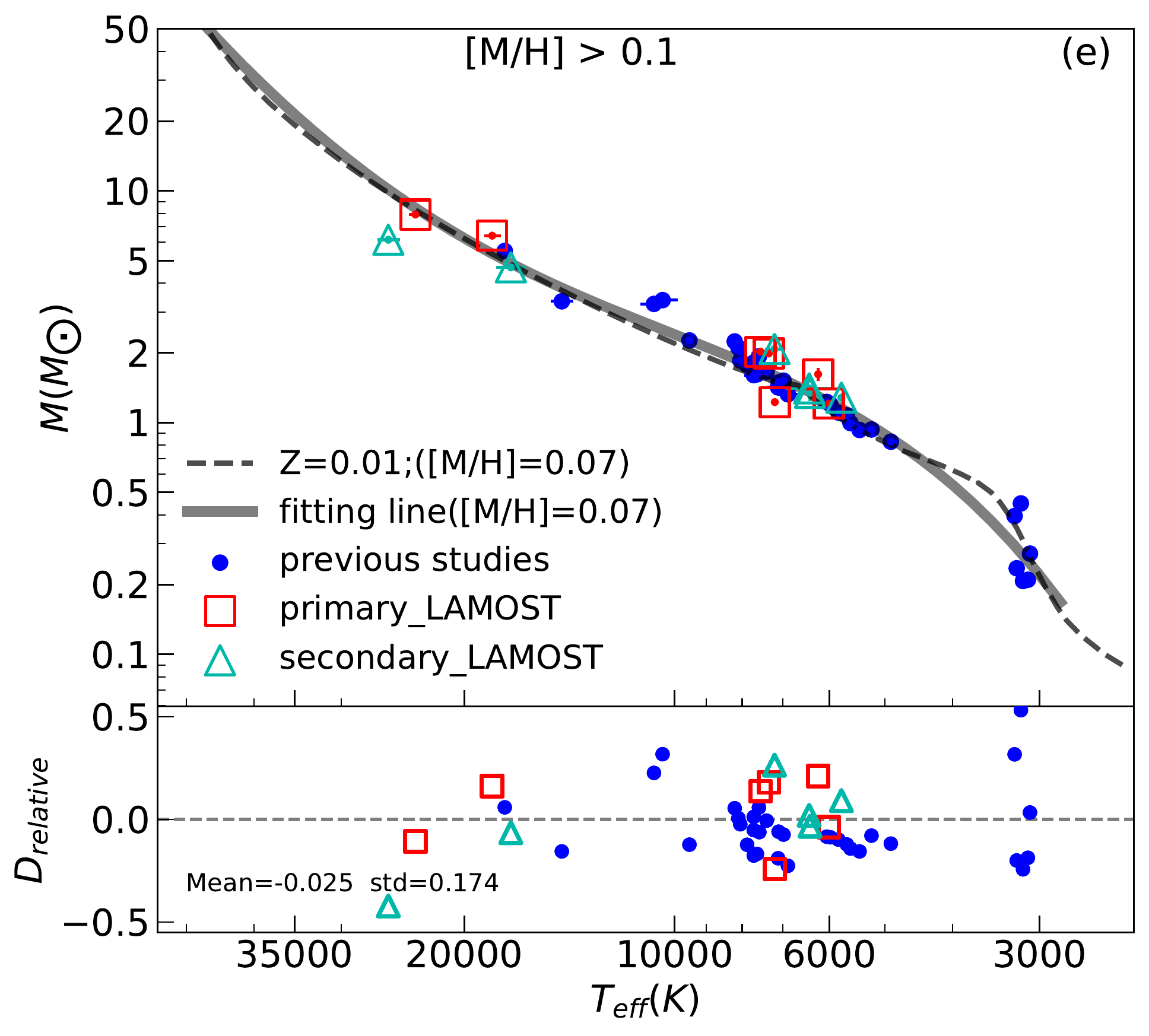}
  }
  \caption{The \teff-\mass\ diagram. Panel (a) shows the total samples with various metallicity. The solid lines stand for the ZAMS with color-coded [M/H] from PARSEC, while the dots are the samples compiled from the literature. The rectangles and triangles are the primary and secondary stars, respectively, from LAMOST MRS. Panels (b)-(e) show the \teff-\mass\ diagrams for the stars with different range of [M/H]. The theoretical ZAMS lines from PARSEC are shown as black dashed lines in the panels. The blue dots, red hollow rectangles, and cyan hollow triangles indicate the stars from literature, the primary and secondary stars from LAMOST MRS, respectively. The parameters of the secondary stars are estimated from PHOEBE. The thick solid gray lines display the best-fit quartic polynomial of \mass\ as a function of \teff\ and \mh. At the bottom of each panels (b)-(e), the relative residuals, defined as $D_{relative}=(M_{obs}-M_{poly})/M_{poly}$ in which $M_{obs}$ and $M_{poly}$ are masses from observation and polynomial model respectively, are illustrated.}\label{fig:fig5}
\end{figure*}

\begin{figure*}[!t]
  \centering
  
  \subfigure{
   \includegraphics[scale=0.38]{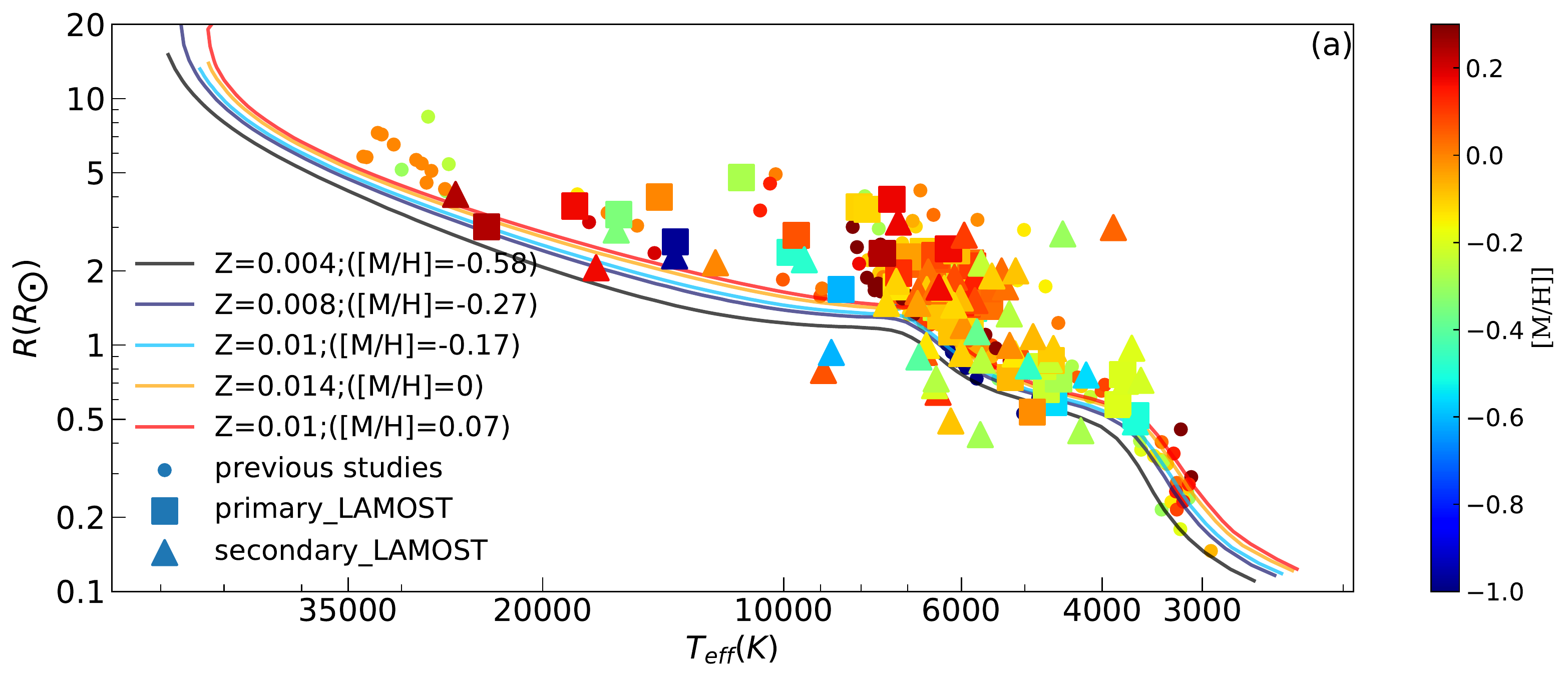}
  }
  \subfigure{
   \includegraphics[scale=0.34]{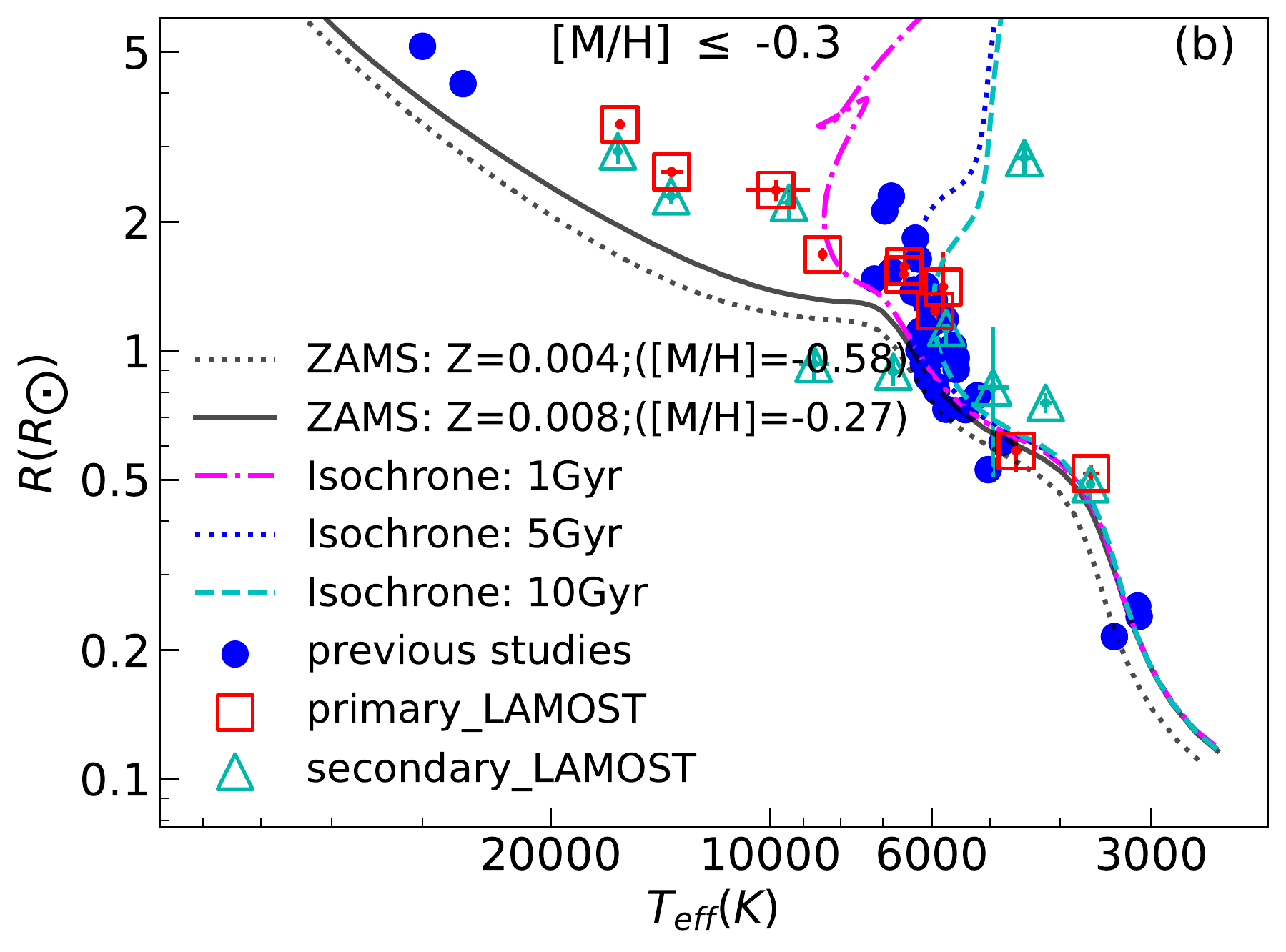}
  }
  \subfigure{
   \includegraphics[scale=0.34]{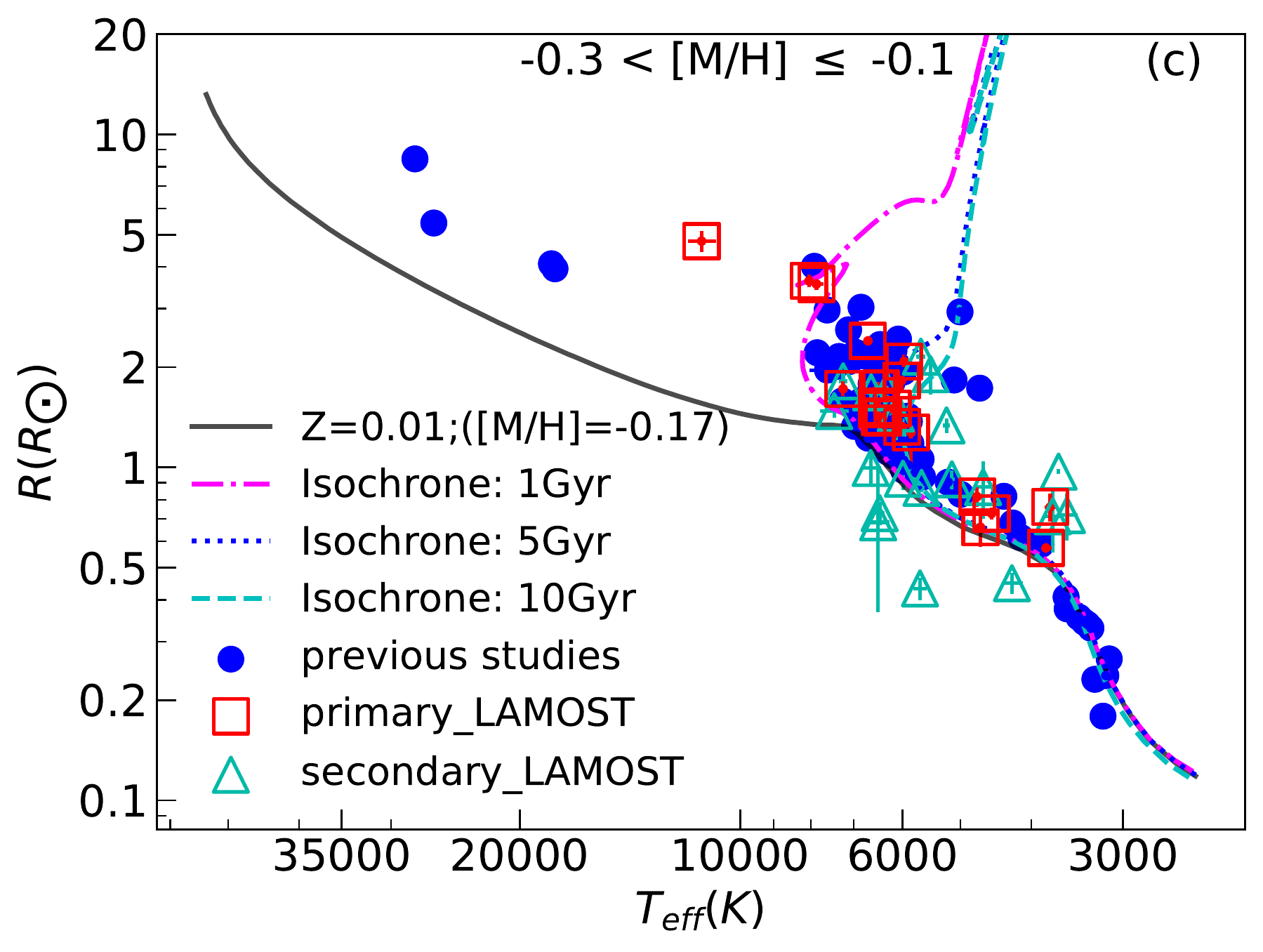}
  }
  \subfigure{
   \includegraphics[scale=0.34]{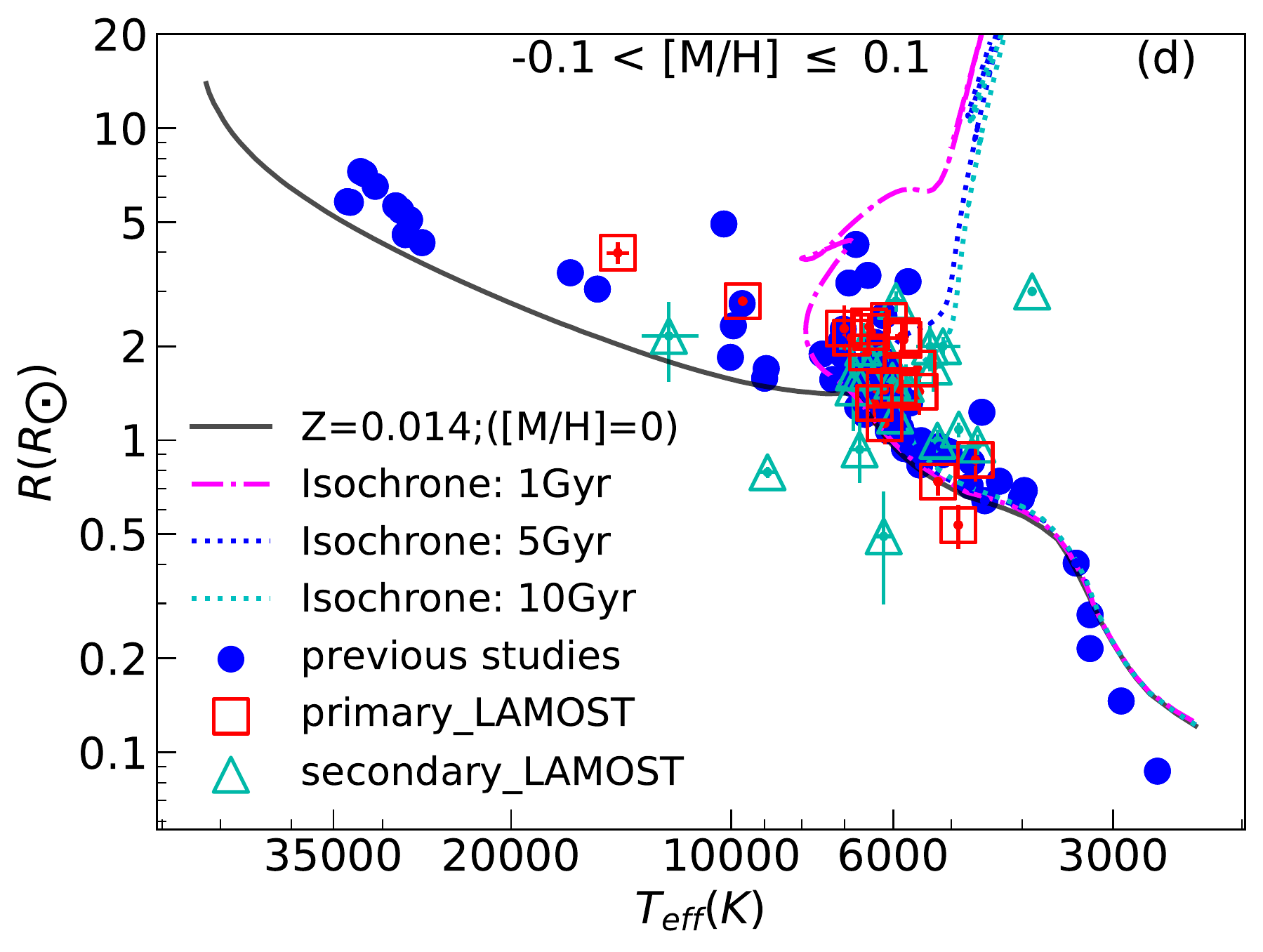}
  }
  \subfigure{
   \includegraphics[scale=0.34]{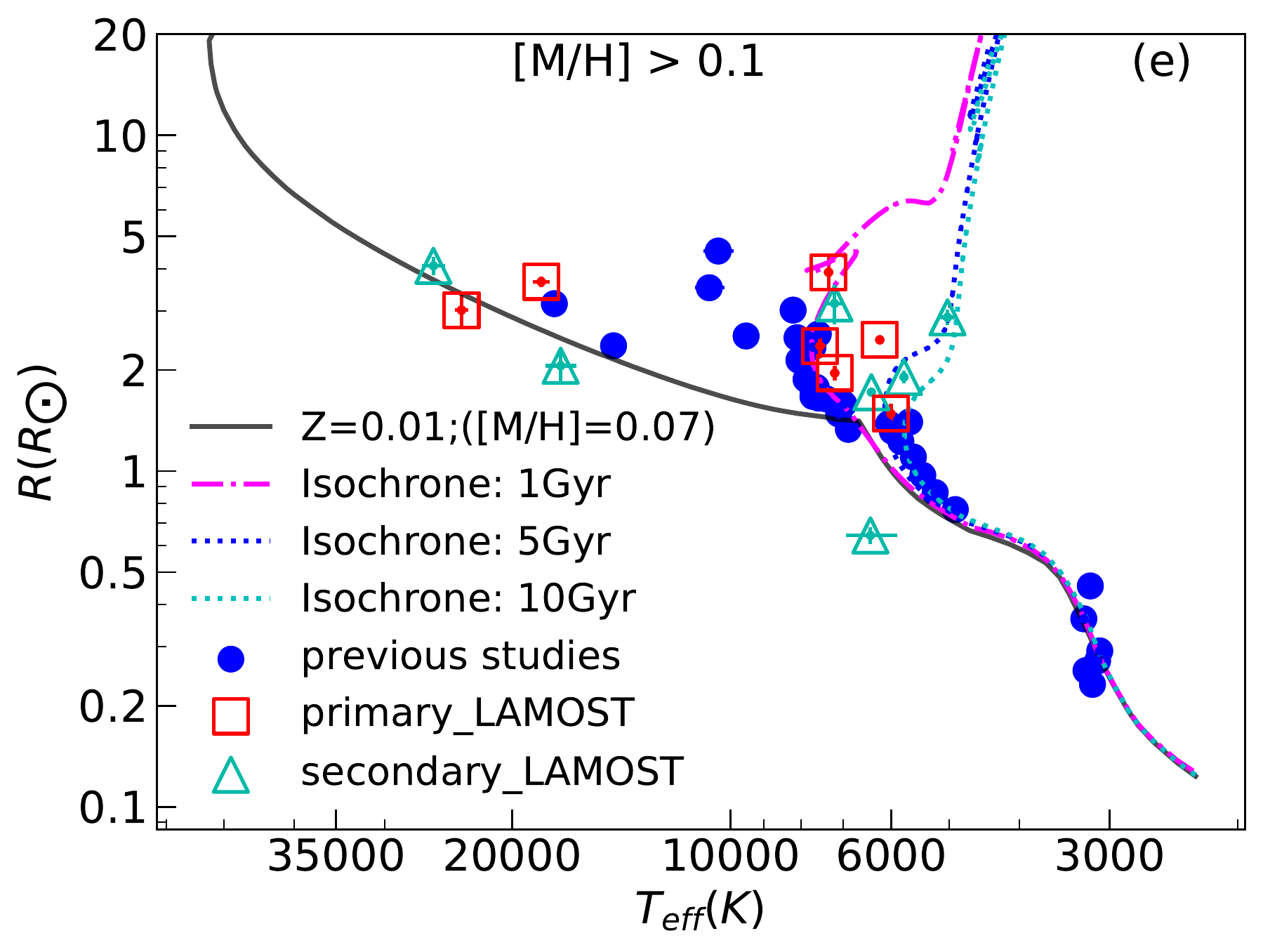}
  }
  
  \caption{The \teff-\radius\ diagram. Panel (a) shows the overall distribution of all samples using dots (known stars from literature), rectangles (primary of the new samples from LAMOST), and triangles (secondary stars from LAMOST). The solid lines indicate the ZAMS lines with color-coded \mh\ from PARSEC. The dots also with color-coded \mh\ are the samples compiled from the literature and the rectangles are the samples selected from LAMOST MRS. Panel (b) shows the \teff-\radius\ diagram with \mh$<-0.3$, in which the blue filled circles, red hollow rectangles, and cyan hollow triangles are same as in Figure~\ref{fig:fig5} (b)-(e). The black solid and dashed lines are the ZAMS lines from PARSEC from $Z=0.008$ and $0.004$, respectively. The magenta dot-dashed, blue dotted and cyan dashed lines stand for three isochrones with age of 1, 5, and 10\,Gyr, respectively, at $Z=0.008$. Panels (c)-(e) are similar to panel (b), but at different metallicities.}\label{fig:fig6}
\end{figure*}

\begin{figure*}[!t]
  \centering
  
  \subfigure{
   \includegraphics[scale=0.38]{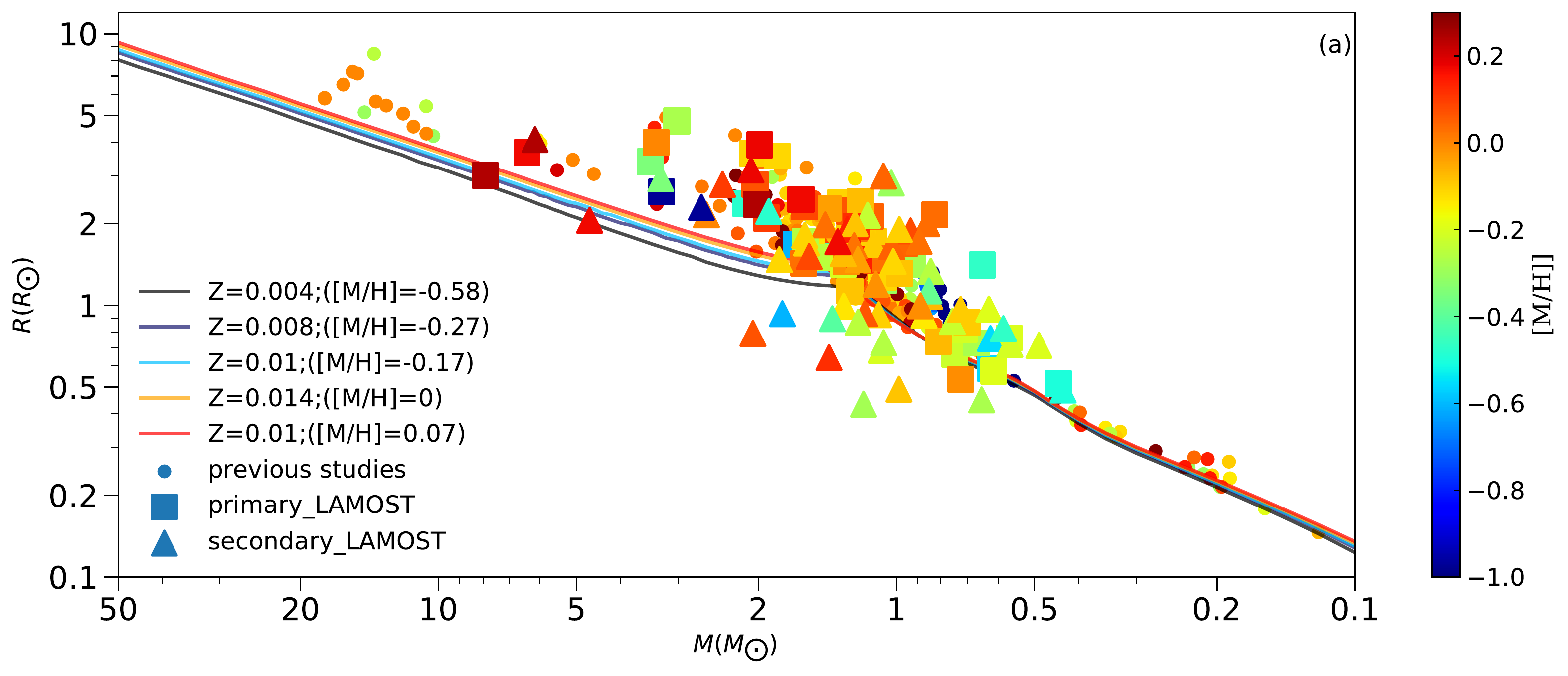}
  }
  \subfigure{
   \includegraphics[scale=0.34]{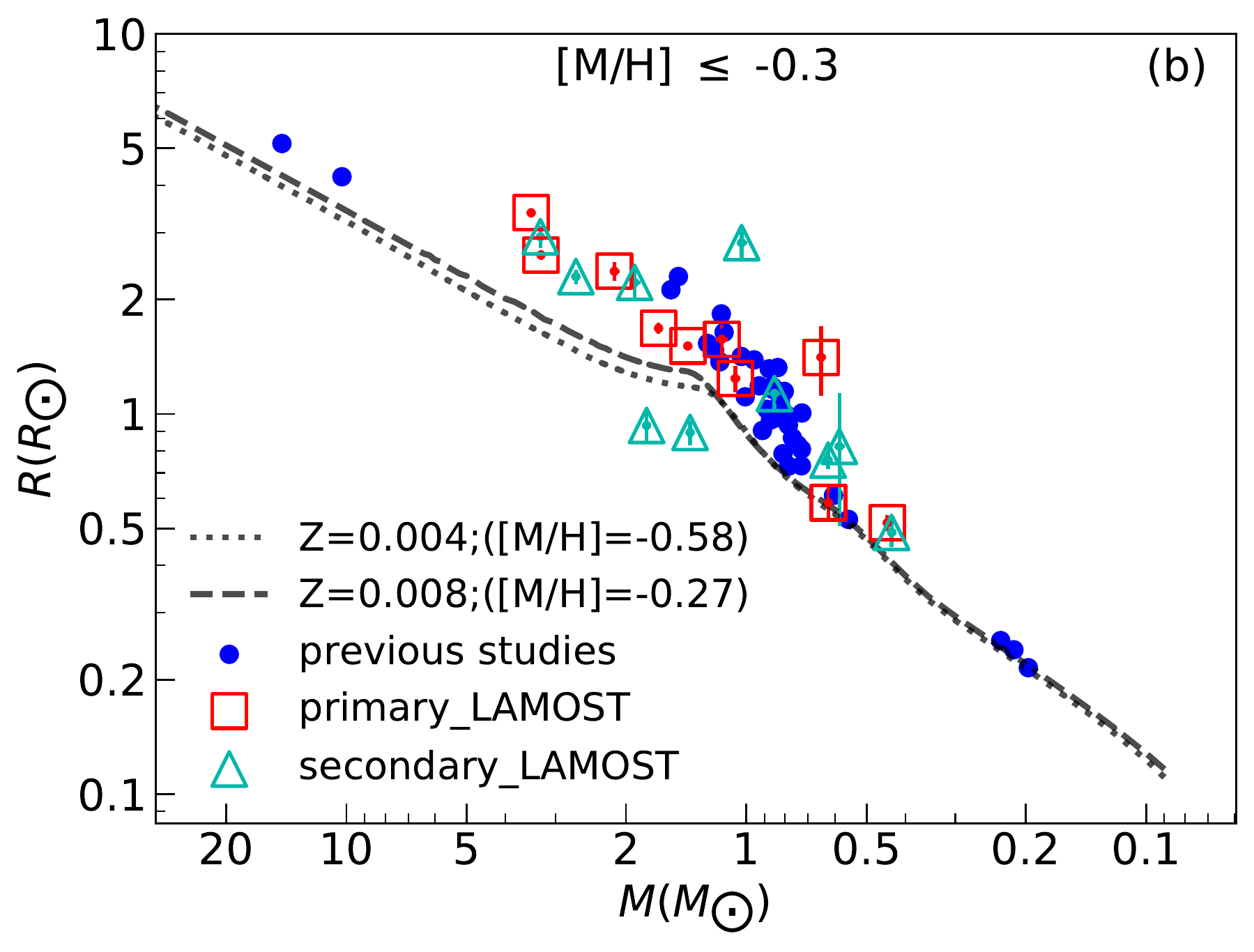}
  }
  \subfigure{
   \includegraphics[scale=0.34]{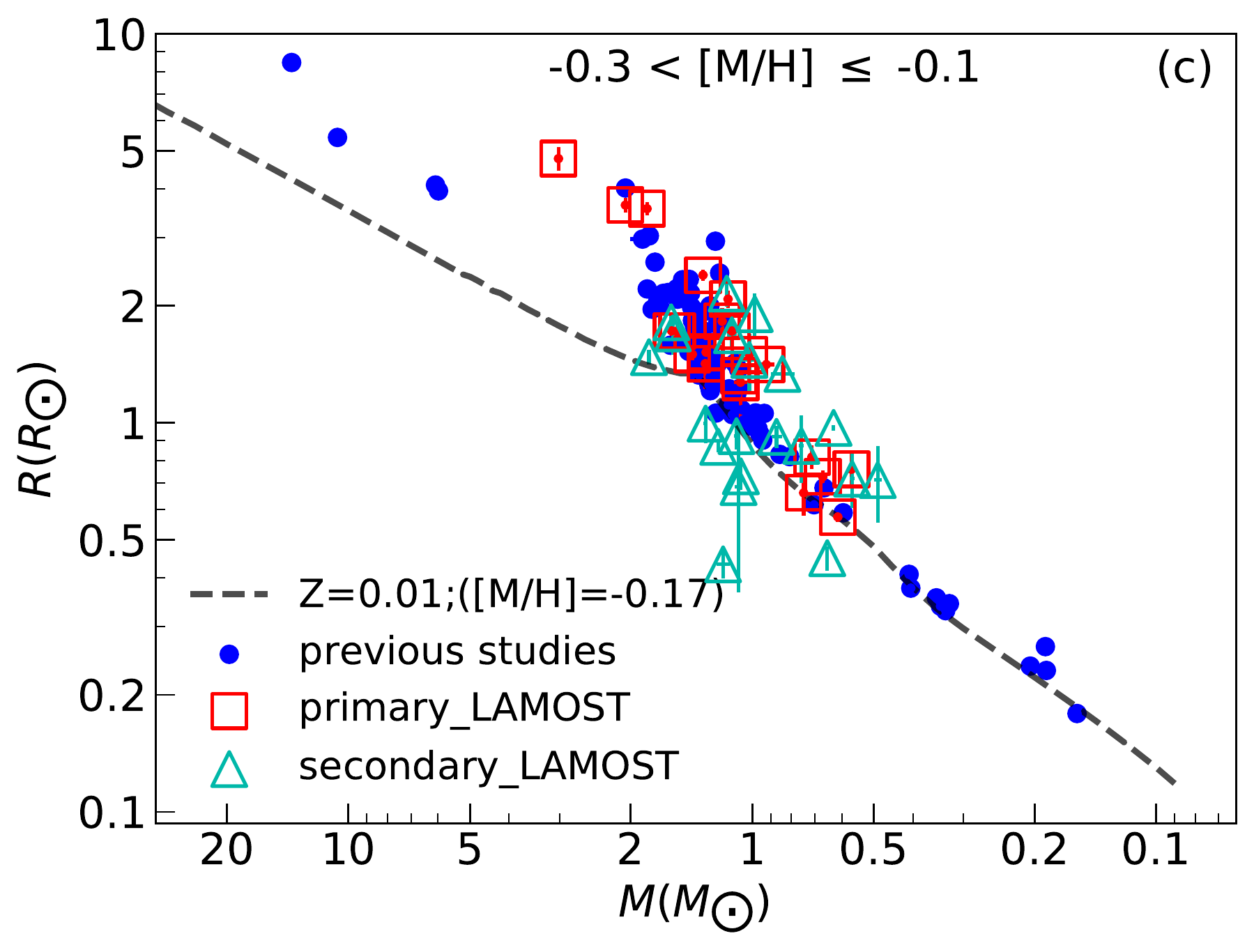}
  }
  \subfigure{
   \includegraphics[scale=0.34]{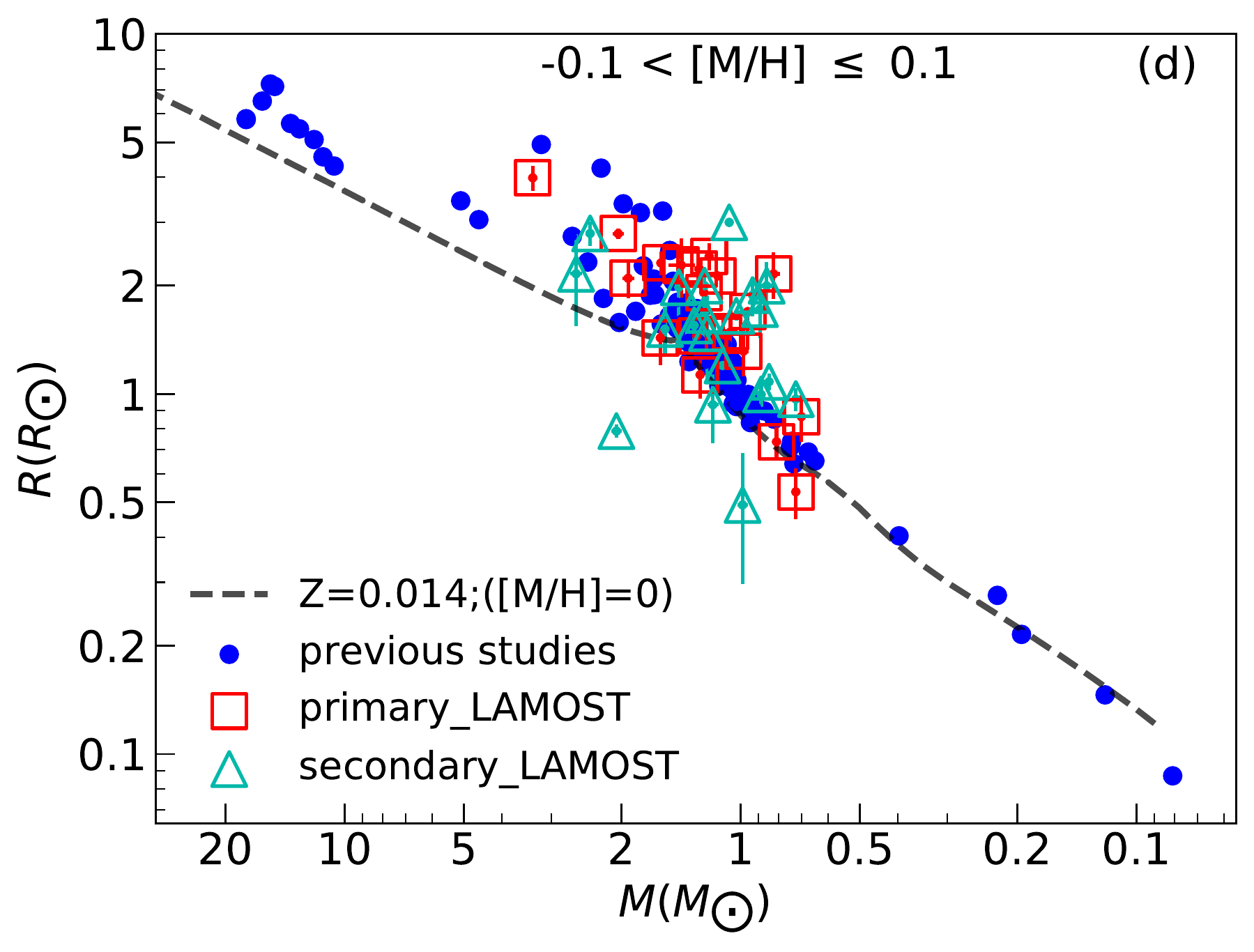}
  }
  \subfigure{
   \includegraphics[scale=0.34]{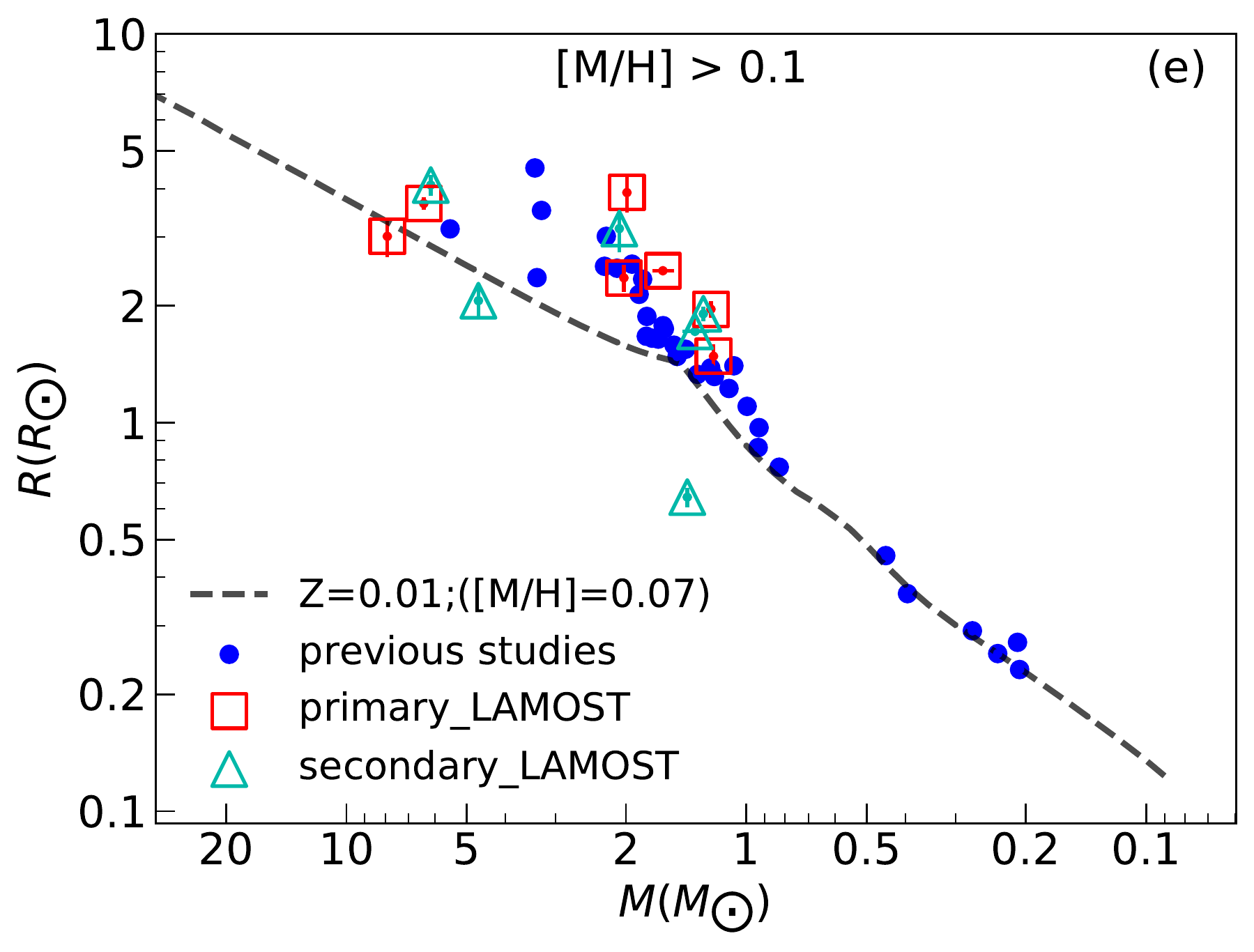}
  }
  \caption{The mass-radius diagram. Panel (a) shows the overall \mass-\radius\ distribution of the samples including both the compiled (dots with color-coded metallicities)) and LAMOST data (rectangles and triangles represent for the primary and secondary stars, respectively). The solid lines with color-coded metallicities are the ZAMS lines with from PARSEC. Panels (b)-(e) display the \mass-\radius\ relations in different ranges of \mh. The dotted or dashed lines indicate ZAMS lines with similar metallicity. The symbols are same as in Figure~\ref{fig:fig5} (b)-(e).}\label{fig:fig7}
\end{figure*}

\subsubsection{\teff-\mass\ relations in different ranges of \mh}

Figure \ref{fig:fig5} shows the \teff-\mass\ relation of the samples. The samples cover a temperature range from 2600 to 38\,000 $K$. In panel (a), the solid lines indicate the zero age main-sequence (ZAMS) with color-coded [M/H] from PARSEC model. Although the model isochrones show metallicity difference in \teff-\mass\ plane, the stellar samples overlapped wuth the isochrones do not show clear gradient in metallicity with \teff$>6000$\,K. This is likely caused by the relatively smaller difference in mass and \teff\ than the measurement uncertainties of the two parameters. However, for the stars with \teff$<6000$\,K, it seems that the metallicity gradient can be seen in the observed samples. The orange and red dots, which represent for high metallicity, are located on top of the bluish symbols, which stand for the low metallicity stars.

\begin{figure*}[!t]
  \centering
  
  \subfigure{
   \includegraphics[scale=0.38]{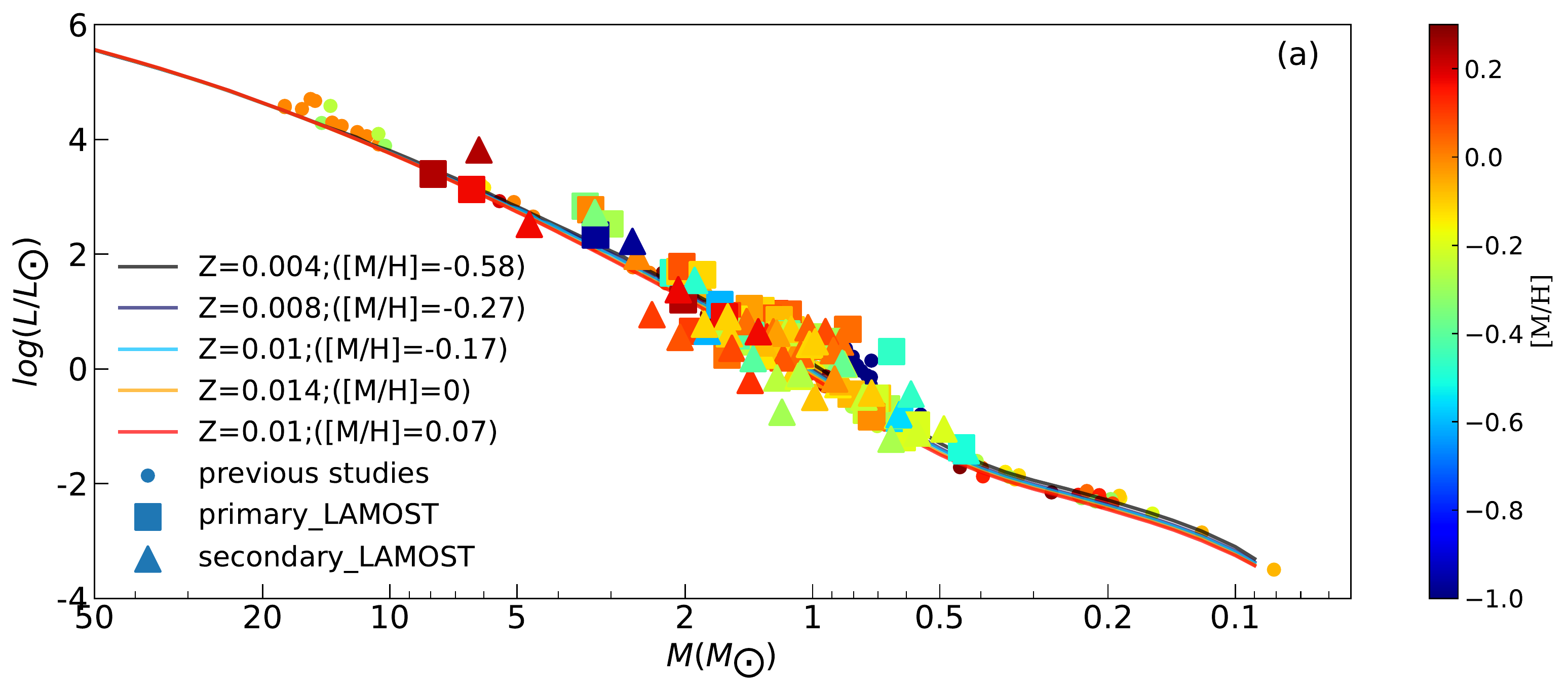}
  }
  \subfigure{
   \includegraphics[scale=0.34]{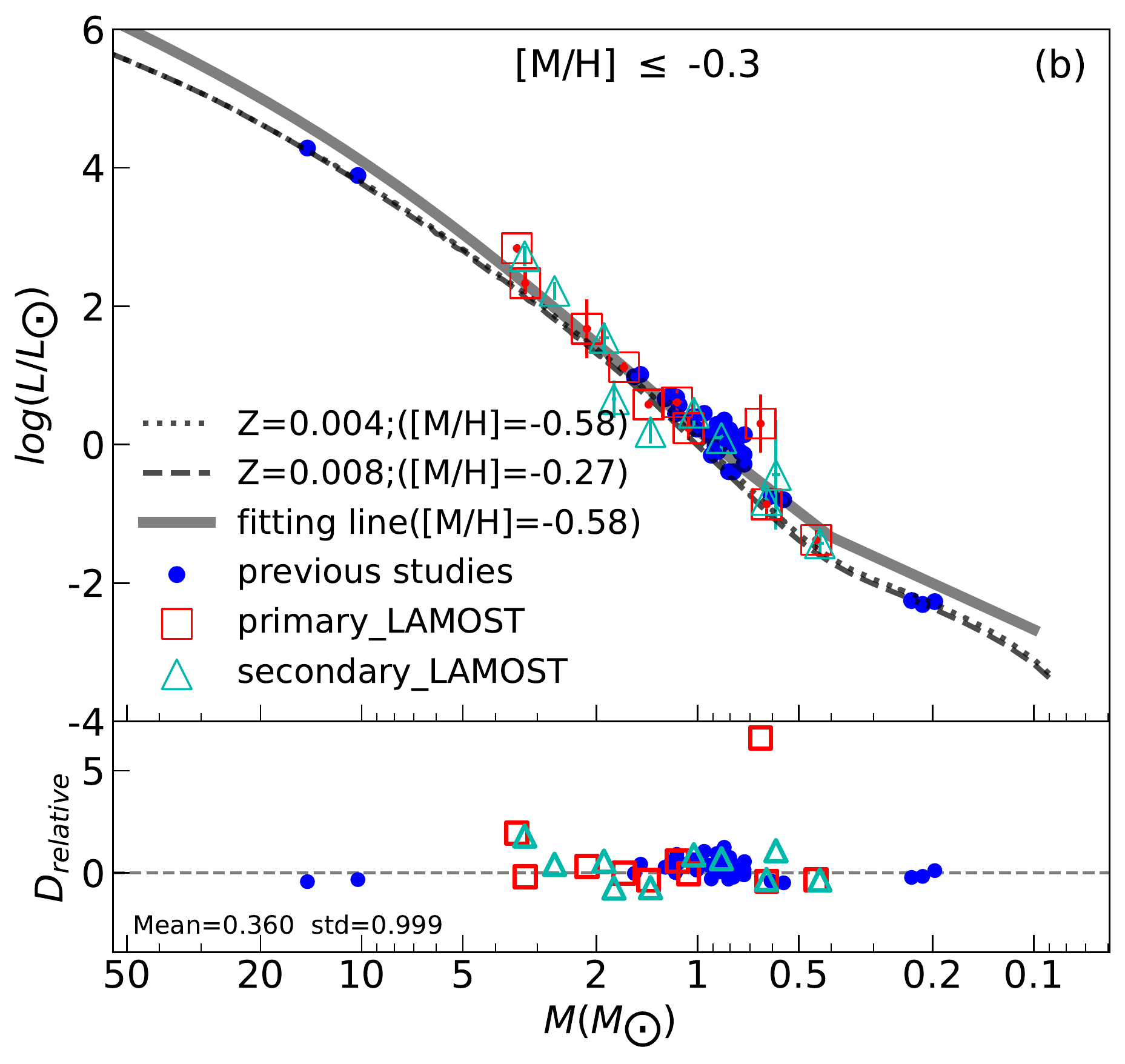}
  }
  \subfigure{
   \includegraphics[scale=0.34]{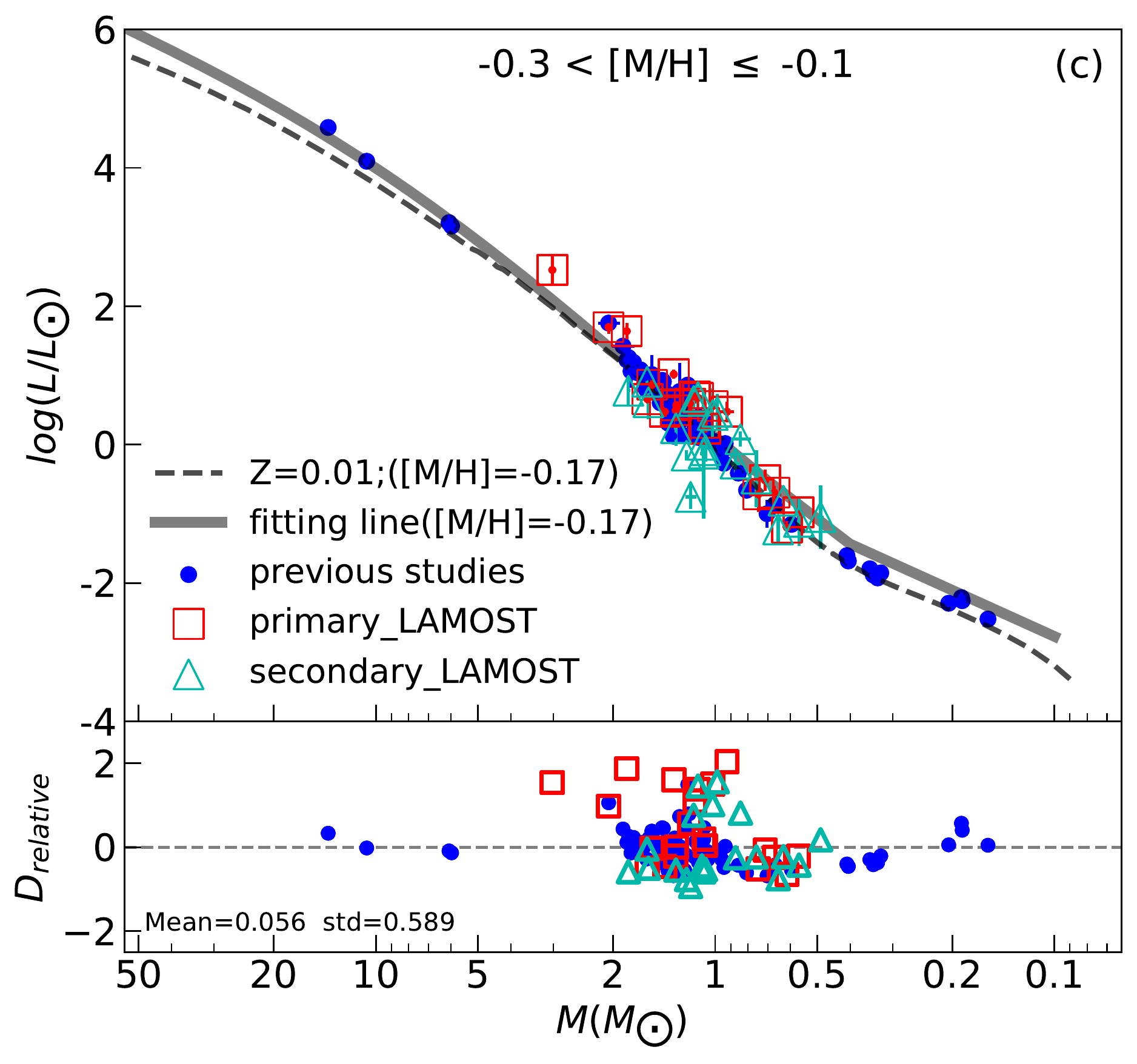}
  }
  \subfigure{
   \includegraphics[scale=0.34]{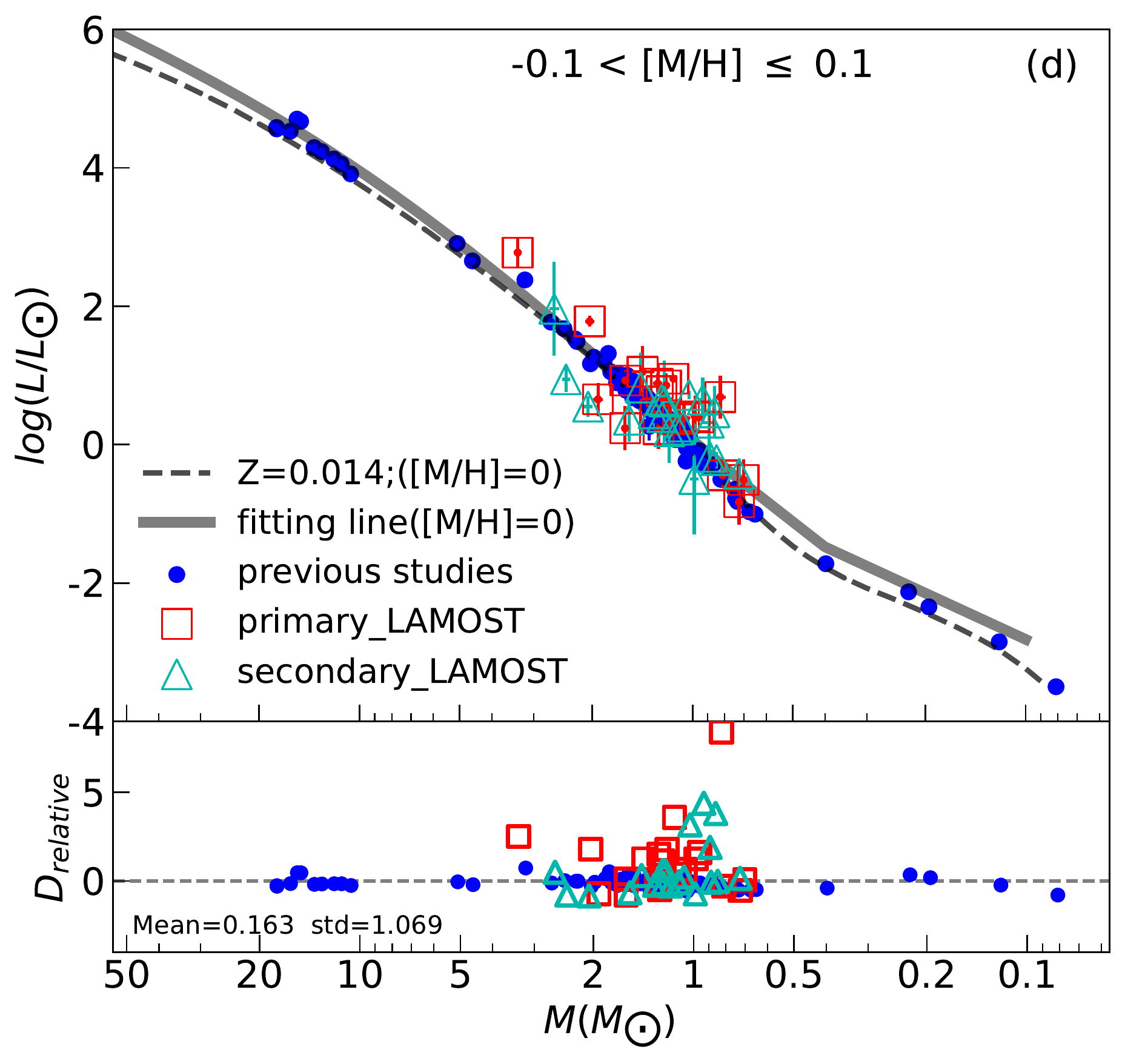}
  }
  \subfigure{
   \includegraphics[scale=0.34]{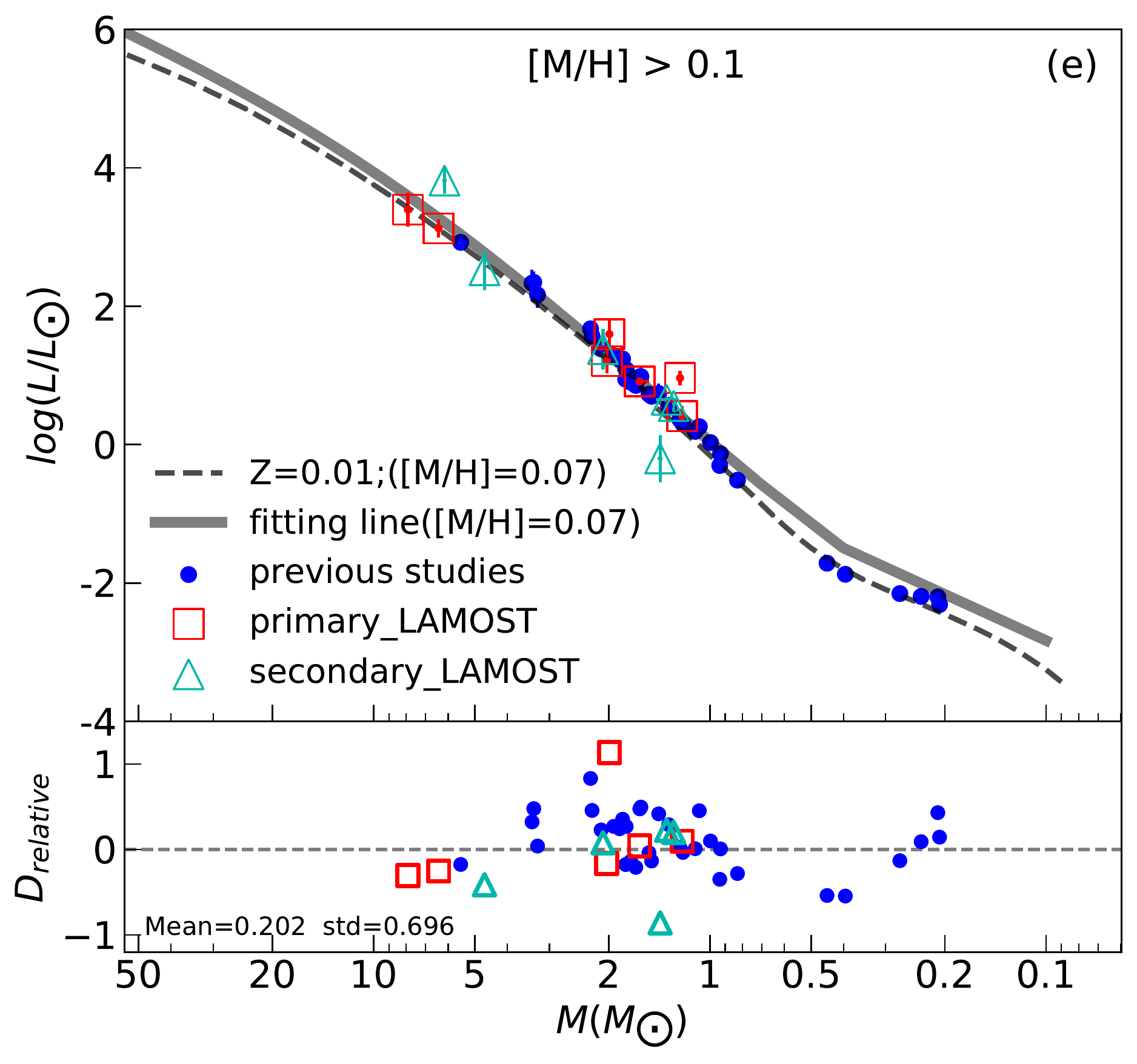}
  }
  \caption{Panel (a) displays the \mass-\lum\ diagram. The solid lines with different colors are the ZAMS lines with color-coded \mh\ from PARSEC. The dots, rectangles, and triangles stand for the stars from literature, the primary and secondary stars from LAMOST MRS, respectively. The colors of the symbols are same as those of the ZAMS lines. Panels (b)-(e) show the\mass-\lum\ diagrams with different range of \mh. The dash lines are the ZAMS lines from PARSEC. The thick solid lines are the best-fit polynomial model. The other symbols are same as in Figure~\ref{fig:fig5} (b)-(e).}\label{fig:fig8}
\end{figure*}

We separately show the \teff-\mass\ relationships at different metallicity bins, as shown from panel (b) to (e), so that detailed comparison between the PARSEC model and the observed stars can be made.

It is seen that most of the stars are well consistent with the stellar models except for the cold stars with \teff$<5000$\,K. The cold, low-mass stars show a steeper slope than the stellar models (dashed lines in the panels) in the variation of the stellar mass when \teff\ declines. This implies that the mass directly estimated from the effective temperature of these cold stars may be over-estimated when \mass$\sim0.1$\,\masssol\ and under-estimated when \mass$\sim0.5$\,\masssol. Although such systematic may only around a few hundredth \masssol, it may potentially flatten the slope of the initial mass function.

Nevertheless, figure~\ref{fig:fig5} implies that the effective temperature combined with a quite coarse estimation of metallicity can approximate the stellar mass in a quite robust way. Therefore, we provide an empirical polynomial model of the stellar mass depending on \teff\ and \mh.
It is not trivial to fit the mass as a function of \teff\ and \mh\ with totally free constraints in the coefficients of a polynomial. Hence, we firstly fit a quartic polynomial to the stellar model and then fix the coefficients of the 3rd and 4th order terms of $\log$\teff\ obtained from the first step and fit the rest coefficients to the data. After a few tests, we adopt the following form of the quartic polynomial
\begin{small}
\begin{equation}\label{eq:RTZ}
\begin{aligned}
    \log(M)=&a_{1}+a_{2}\log(T)+a_{3}\log(T)^{2}+18.398\log(T)^{3}\\
    &-0.998\log(T)^{4}+a_{4}Z,
\end{aligned}
\end{equation}
\end{small}

where $T$ and $Z$ represent for \teff\ and \mh . The best-fit $a_{1}$, $a_{2}$, $a_{3}$, $a_{4}$ are 414.761, 372.928, -124.902, 12.088, respectively. It is noted that the relationship derived among these various parameters only apply for unevolved stars.
The gray solid lines in the panels (b)-(e) indicate the best-fit polynomial model. The relative residuals showing at the bottom of panels (b)-(e) are around 11$\sim$18\%.

In panel (b), benefited from the observation of lower metallicity stars in the outer disk of the Milky Way by LAMOST, we can see that there is an extension at the part of high \teff. 

In panels (c) and (d), which correspond to $-0.3<$\mh$\leq-0.1$ and $-0.1<$\mh$\leq+0.1$, the numbers of samples from LAMOST fill some vacancies in \teff\ so that the coverage of \teff\ becomes more continuous. In panel (e), for the stars with \mh$>+0.1$, the LAMOST samples also extend the range of \teff\ to higher value.

\subsubsection{\teff-\radius\ relations in different ranges of \mh}

Figure \ref{fig:fig6} shows the \teff-\radius\ relation of the samples. Similar to figure~\ref{fig:fig5}, we draw the relation for the whole stars in panel (a) and those for different metallicity ranges in the rest four panels.
Unlike \teff-\mass\ relation, \teff-\radius\ relation depends on the age of stars. Therefore, the stars in panel (a) are mostly located above the ZAMS lines. The gradient of metallicity in the \teff-\radius\ plane is not seen in the stellar samples since that \radius\ is either affected by age or by uncertainties of the \radius\ and \teff\ estimates.

In panels (b)-(e), the \teff-\radius\ relations with different ranges of [M/H] are displayed, respectively. The solid lines indicate the ZAMS lines at different [M/H], while dot-dashed, dotted, and dashed lines with color-coded ages are the isochrones with the ages of 1, 5, and 10 Gyr, respectively. The influence of age, which moves older stars upward out of ZAMS, especially near \teff$\sim6000$\,K, is clearly seen in all the panels. Comparing to the theoretical isochrones, it implies that, with accurate measurement of radius and effective temperature, one would also determine the age of stars, not only for turn-off stars (which is one of the most sensitive kinds), but also for stars at a large range of \teff.

In panel (d), for the stars with \teff\ $<$ 4000 $K$, as seen in figure \ref{fig:fig4} (c), a clear deviation between the observed stars and the theoretical ZAMS is found. This may be caused by the inaccurate atmospheric model for cool stars.

\subsubsection{\mass-\radius\ relations in different ranges of \mh}

Figure \ref{fig:fig7} shows the \mass-\radius\ relations of our samples including both compiled and LAMOST MRS data. In principle, the trend is quite consistent with that presented by \citet{2018MNRAS.479.5491E}.
In panel (a), the ZAMS lines show that the difference of \mass-\radius\ relations in different \mh\ is small, at least in the range of $-0.58<$\mh$<+0.07$. In particular, when \mass$<0.5$, the ZAMS lines with different \mh\ are almost overlapped with the observed data.

Panels (b)-(e) separately show the \mass-\radius\ relations in different ranges of \mh. Compared to the \teff-\radius\ relations, they display that the \mass-\radius\ relation of the observed samples are well consistent with the stellar models, especially at \mass$<1$\,\masssol, in different ranges of [M/H].

Almost all the data samples are located on above the theoretical ZAMS lines. Although the stellar masses does not significantly change from ZAMS, their radii enlarge to the extent that we are able to measure. In particular, for the stars with mass around 2\,\masssol, the radii can extend to a factor of a few from the initial value at ZAMS with same mass. The explanations of the phenomenon are mainly considered as the evolution of stars~\citep{2010AARv..18...67T}, ages, rotation~\citep{2011ApJ...742..123I,2011ApJ...728...48K}, and magnetic fields~\citep{1986A&A...166..167S,2012ApJ...757...42F,2013ApJ...765..126M}.

\subsubsection{\mass-\lum\ diagrams in different ranges of \mh}

The luminosity, which is defined as $L=4\pi R^{2}\cdot \sigma T_{eff}^{4}$, is derived from \teff\ and \radius. As logarithmic form, $\log L=4\log T_{\rm eff}+2\log R-15.045$, in which \lum, \radius\ are in units of the Sun. 

Figure \ref{fig:fig8} shows the \mass-\lum\ diagram. In panel (a), the color-coded solid lines are the ZAMS lines with different \mh, the dots are stars from literature, and the rectangles are those from LAMOST MRS. Colors of these symbols represent for the metallicity with same code as the ZAMS lines. 
It is seen that \mh\ is almost independent in the \mass-\lum\ relation, which is consistent with the stellar model showing in ZAMS lines.
Furthermore, at around 1\,\masssol, a larger dispersion in observed data can be clearly seen in the panel. This is likely due to the influence of age.
 
The \mass-\lum\ diagrams with different \mh\ are shown in panels (b)-(e).  It shows that, at each metallicity bin, the observed data are all well consistent with the stellar models. It is noted that at low mass end with \mass$<1$\,\masssol, unlike the \teff-\mass\ and \teff-\radius\ relations, the data also well agree with the models. This means that, although \radius\ is not well predicted by stellar model for low mass stars, the luminosity predicted by model seems quite good.

We also conducted an polynomial model of \mass-\lum-\mh\ relation and find the best-fit coefficients of the model using the stellar samples. The following relationship also only apply for the unevolved stars. The polynomial model is written as: 


\begin{small}
\begin{equation}\label{eq:MLZ}
\begin{aligned}
     \log(L)=&a_{1}+a_{2}\log(M)+a_{3}\log(M)^{2}\\
    &-0.755\log(M)^{3}+0.189\log(M)^{4}+a_{4}Z,
\end{aligned}
\end{equation}
\end{small}
in which, $a_{1}$, $a_{2}$, $a_{3}$, $a_{4}$ are 0.066, 4.141, 0.314, -0.245, respectively. The coeffecients of the 3rd and 4th order terms of $\log M$ are determined from the polynomial fitting to the model. The best-fit \mass-\lum\ polynomial model is shown as thick solid lines for different \mh\ in panels (b)-(e), respectively. The relative difference between observed data and model ($L_{fit}$), $D_{relative}=(L-L_{fit})/L_{fit}$, is shown in the bottom of each panel.

\begin{figure*}[!t]
  \centering
    \subfigure{
   \includegraphics[scale=0.15]{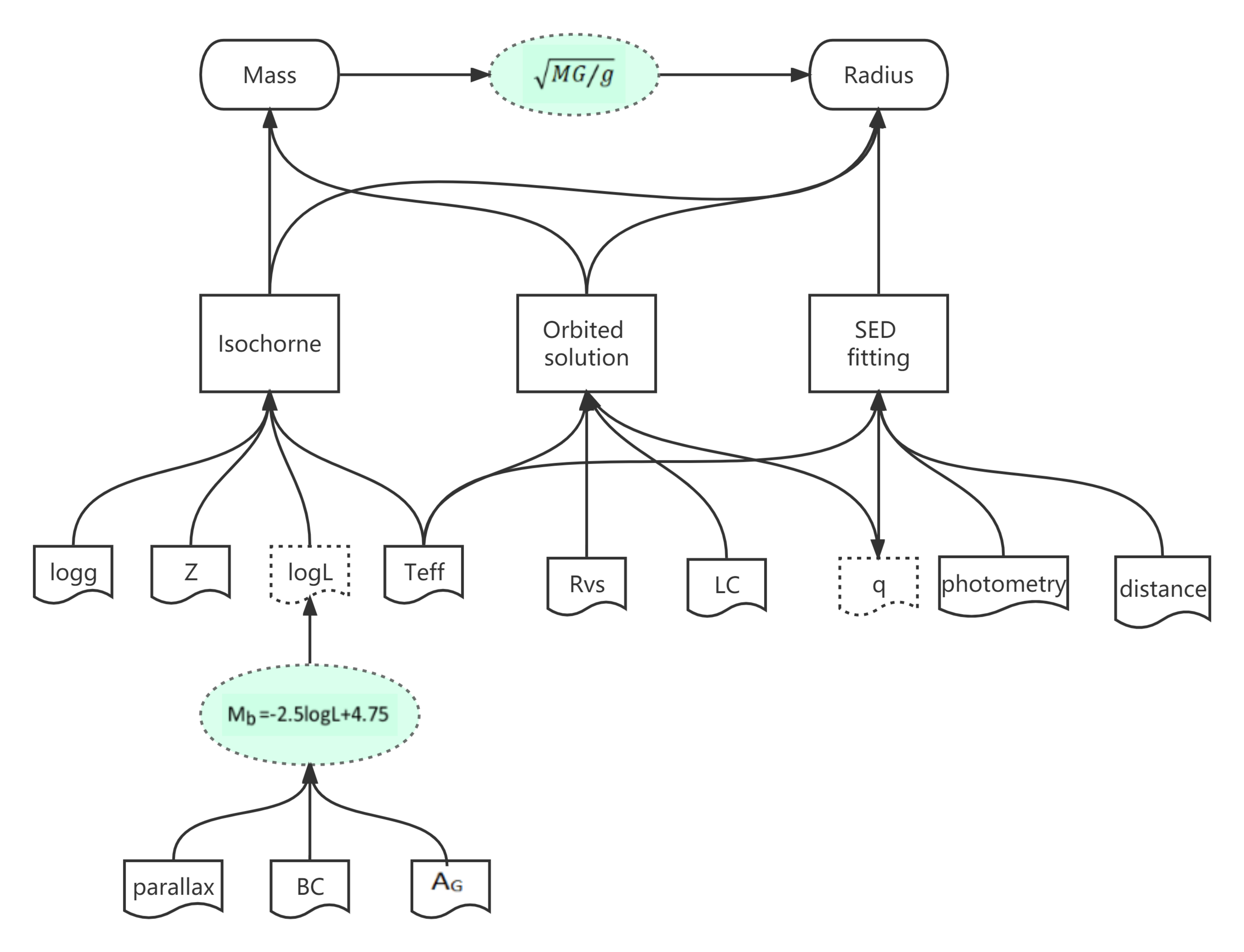}
  }
  \caption{Summary of parameter measurement approaches. The rectangles in the second layer are measurements for mass and radius. In the third layer, the solid blocks are the parameters directly obtained from observation and the dotted blocks are the parameters derived from some relations. The green circles are relationships between parameters.}\label{fig:relation}
\end{figure*}

\section{Discussion} \label{sec:Discussion}

The high-precision, model-independent mass and radius derived from binaries allow us to calibrate different methods of estimation of mass and radius for single stars from observational atmospheric parameters (e.g. \teff, $\log{g}$ and [M/H]).

In this section, we compare the \mass\ and \radius\ derived from orbital solutions with those from other measurements. 

There are lots of approaches to obtain stellar mass and radius. Figure~\ref{fig:relation} summarizes the usual paths to the two parameters.

In principle, there are three kinds of approaches, which are drawn as three rectangles in the second layer, to estimate \mass\ and \radius. 

Each approach requires observed parameters drawn in the third layer in figure~\ref{fig:relation}. The solid blocks indicate the parameters directly obtained from observation and the dotted blocks are the parameters derived from some known relationships, which are shown in the green circles with dotted lines.

Other than orbital solution, the stellar mass can also be determined using isochrone fitting with \teff, \logg, and metallicity as inputs. Luminosity can also be used in the isochrone fitting for the stellar mass. In total, there are three usual methods for mass estimation: orbital solution, isochrone fitting with \teff, \logg, and $Z$, and isochrone fitting with additional $\log L$ derived either from parallax. These 3 methods are listed in Table~\ref{tab:relation}.

To estimate the stellar radius, one usually have four other methods than the orbital solution. The first two use isochrone fitting with or without luminosity. The luminosity required in radius estimation should be determined from parallax. The other two methods are based on SED fitting either with multi-band photometry, \teff, and distance or with additional mass ratio, $q$ from other methods.

Table~\ref{tab:relation} summarizes all these methods with different input parameters and the results of the comparison with binary orbital solution. We compare the results of mass and radius from various methods and assess the performance of them based on the results from the orbital solutions.

\begin{table*}[t]
\scriptsize
\caption{Summary of the usual mass and radius measurement methods.}\label{tab:relation}
\centering
\begin{tabular}{c|c|c|c|c|c}
\hline
\hline
  \multicolumn{1}{c|}{Parameters} &
  \multicolumn{1}{c|}{Definition} &
  \multicolumn{1}{c|}{Method} &
  \multicolumn{1}{c|}{Input Parameters}&
  \multicolumn{1}{c|}{Mean relative difference\footnote{relative difference: (M - $M_{X}$ )/$M_{X}$, $X$ are various methods}}&
  \multicolumn{1}{c}{Random error}\\
\hline
& M & Orbited solution & RVs(RV$_{1}$ and RV$_{2}$), LC, \teff(primary star);&-&- \\
\cline{2-6}
& $M_{g}$ & Isochorne fitting & \teff,\logg, \mh &0.010&0.143\\
\cline{2-6}
& $M_{LG}$ & Isochorne fitting& \teff,\logg, \mh\ ; logL;&-0.289&0.188 \\
\hline
& R & Orbited solution & RVs(RV$_{1}$ and RV$_{2}$), LC, \teff(primary star); &-&-\\
\cline{2-6}
& $R_{g}$ & Isochorne fitting& \teff,\logg, [M/H]; &0.011&0.164\\
\cline{2-6}
 Radius& $R_{LG}$ & Isochorne fitting& \teff,\logg , [M/H], log L; &-0.322&0.191  \\
\cline{2-6}
& $R_{SED}$ & SED fitting & \teff($T_{1}$ and $T_{2}$), q, distance and photometry from Gaia; &-0.029&0.205\\
\cline{2-6}
& $R_{Mg}$ & $\sqrt{MG/g}$ & mass,\logg\ ; &0.027&0.292\\
\hline
\end{tabular}
\end{table*}

\begin{figure*}[!t]
	\centering
	
	\subfigure{
		\includegraphics[scale=0.4]{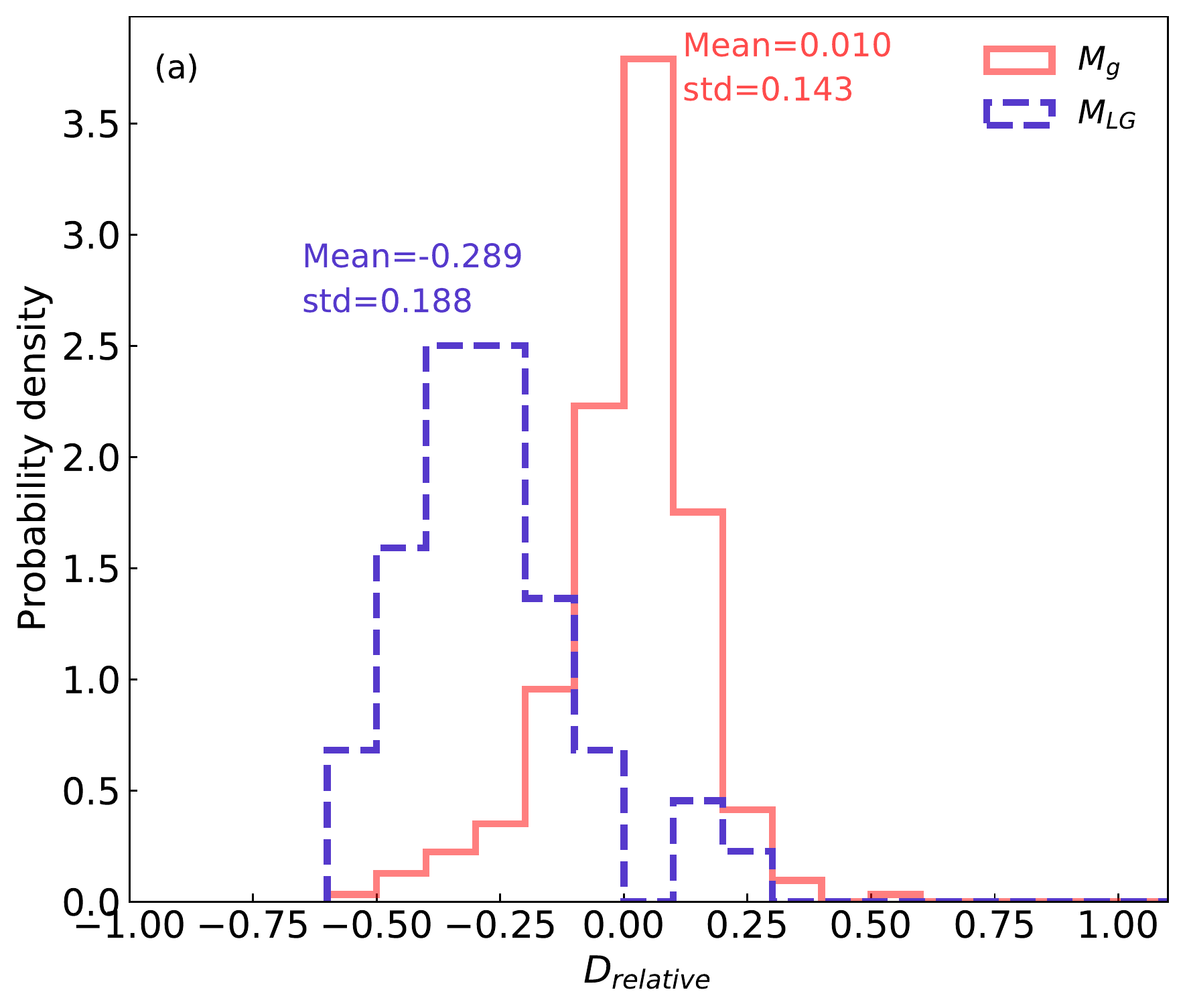}
	}
	\subfigure{
		\includegraphics[scale=0.4]{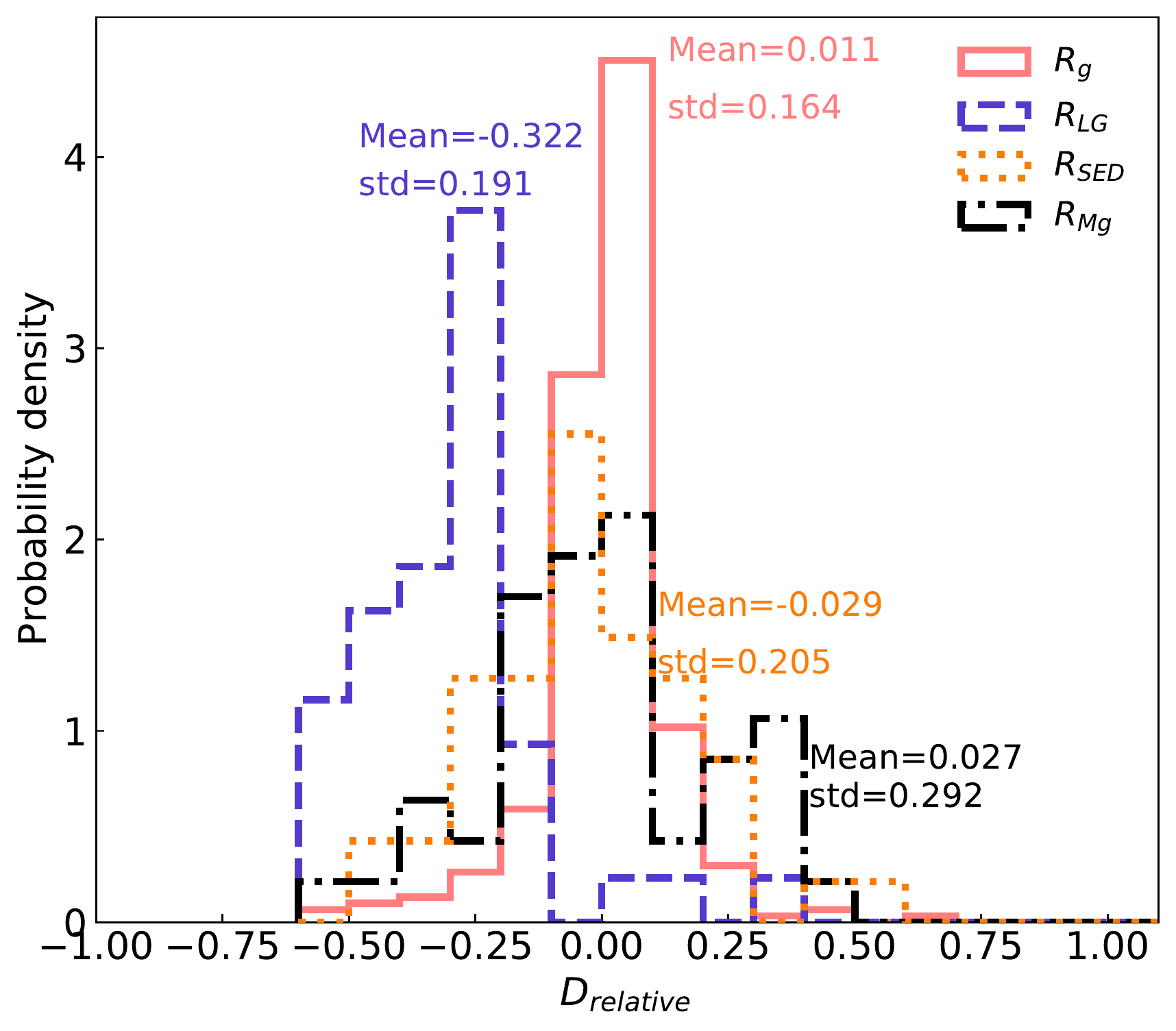}
	}
	
	\caption{Panel (a) shows the histogram of the relative residuals ($D_{relative}=(M_x-M)/M$, where $x$ represents either $g$ or $LG$) between dynamic mass (\mass) and the mass derived from model or relations. Panel (b) shows the histogram of the relative residuals ($D_{relative}=(R_x-R)/R$, where $x$ stands for $g$, $LG$, $SED$, or $Mg$) between dynamic radius (\radius) and the radius derived from model or relations.}\label{fig:fig10}
\end{figure*}

In general, all isochrone fitting approaches have similar process, that is, to find the best-fit physical parameters in synthetic stellar models with most similar observed atmospheric parameters and/or luminosity.

In this study, Nearest Neighbor Search (NNS) is applied to find the best-match stellar atmospheric parameters and/or luminosity in the PARSEC \citep{2012MNRAS.427..127B} theoretical stellar model grid. The atmospheric parameters and/or luminosity to be matched are defined as a vector ${\mathbf{\theta}}=$(\teff, \logg, \mh ) or (\teff, \logg, \mh, \lum), in which \lum\ is derived from parallax. The PARSEC isochrone grid with \mh\ in (-2.0 dex $\sim$ +0.5 dex) and ages in (10\,Myr $\sim$ 10\,Gyr) sets up the search space S=\{$s_{1}$, $s_{2}$,$s_{3}$......$s_{n}$\} with various combination of \teff, \logg, and \mh. Firstly, we generate a searching point (${\theta_{th}}$) following a uniform distribution within the range of $\pm$3 times measurement errors for ${\theta}$. Second, the distance between PARSEC isochrone grid and ${\theta_{th}}$ is calculated such that
\begin{equation}\label{eq:NN}
	d({\theta_{th,i}},s_{n,i})=\sqrt{\sum_{i=1}^m((s_{n,i}-{\mathbf\theta_{th,i}})/C_{i})^2},
\end{equation}
where $i$ represents \teff, \logg, \mh \, or with additional \lum\ and $m=3$ or $4$ depending whether luminosity is among the input parameters. $C$ is a custom coefficient to unify the parameters scale. We empirically adopt $C$ as 1000 for \teff, 0.25 for \logg\ and \lum, and 0.1 for \mh. Third, we search for the closest point to ${\theta}$ in S and adopt the mass and radius of the nearest neighbor point as the best estimates. Finally, we obtain the mass and radius estimation with measurement error by repeating the above steps 5000 times. In each time, the input $\theta$ are randomly drawn from a Gaussian distribution adopting the uncertainties of $\theta$ as scales.
During this process, \teff, \logg, and \mh\ are obtained from the observed spectra. \lum\ is measured using the bolometric magnitude, which is 
\begin{equation}
		\log L=0.45(M_{b,\odot}-M_{b}),
\end{equation}
where $M_{b,\odot}$ is the absolute bolometric magnitude of the Sun. The absolute bolometric magnitude $M_{b}$ of each star is derived from
\begin{equation}
	 	M_b = M'_G-A_G+BC_G,
\end{equation}
where $M'_G$ is derived by the following equation \citep{2018A&A...616A..10G}:
\begin{equation}\label{eq:mg}
	 M'_G=G+5+5log_{10}(\varpi/1000)
\end{equation}
based on photometric and astrometric data of Gaia EDR3 \citep{2020arXiv201201533G}. In Eq.~\ref{eq:mg}, $G$ is G magnitude, $\varpi$ is the parallax in milli-arcsecond. The extinction $A_{\rm V}$ is given from the 3-D dust map provided by \textit{dustmaps bayestar} \citep{2019ApJ...887...93G}. $A_{\rm G}$ is further derived from $A_{\rm V}$ using the extinction coefficient from \cite{2019ApJ...877..116W}. Finally, the bolometric correction of stars are obtained from \cite{2019A&A...632A.105C}.
	  
In figure~\ref{fig:fig10}, panel (a) shows the histogram of the relative residuals between dynamical mass ($M$) and other mass estimates. And panel (b) displays the histogram of the relative residuals between dynamical radius and other radius estimates.

It is seen from the red line that, with \teff, \logg, and \mh, the stellar mass estimated from isochrone fitting can reach random error of about 14.3\%. The systematic bias in $M_{g}$ is only 1.0\%, which is quite small. This implies that the stellar model applied in the isochrone is quite accurate compared to the dynamical mass.

The blue dashed line shows the performance of $M_{LG}$ estimates, which is derived from isochrone fitting of the LAMOST MRS samples with ${\rm RUWE}<$1.5. The luminosities are estimated based on \emph{Gaia} parallax. The relative systematic bias of $M_{LG}$ is -28.9\%, which is slightly larger compared to the random error, which is 18.8\%. 

It seems that the isochrone fitting without luminosity can well reproduce stellar mass, while with luminosity, it tends to overestimate stellar mass. The systematic bias and larger random error of $M_{LG}$ is probably due to the large uncertainties of bolometric correction in $G$-band.

Panel (b) shows the difference between dynamical radius ($R$) and the those derived from other 4 methods. The red solid line shows that the isochrone derived radius $R_g$ using \teff, \logg, and \mh\ as inputs tightly follows the dynamical radius. The systematic difference between $R_g$ and $R$ is only 1.1\% with random error of 16.4\%. 

The blue dashed line shows the comparison between dynamical $R$ and $R_{LG}$, which is obtained from isochrone fitting using \teff, \logg, \mh, and \lum. Similar to the mass estimate $M_{LG}$, a significant overestimation of 32.2\% occurs in $R_{LG}$. This is again likely cause by the larger uncertainty of bolometric correction.

The orange dotted line shows that the SED derived radius $R_{SED}$ is quite consistent with the dynamical values. The mean difference is only -2.9\% with random error of 20.5\%.  The surface gravity-derived radius $R_{Mg}$ shows similar performance as seen in the black dot-dashed line. The bias is 2.7\% and the random error is 29.2\%, slightly larger than that for $R_{SED}$. Note that $R_{SED}$ adopts accurate distance derived from parallax given by \emph{Gaia} and $R_{Mg}$ uses dynamical mass in the calculation. These accurate parameters are helpful to constrain the accuracy of $R_{SED}$ and $R_{Mg}$.

The accuracy of atmospheric stellar parameters \teff, \logg, [M/H], and \lum\ are critical in the precision of mass and radii estimation from the stellar model. If these parameters can not be measured in high-precision, the mass and radius can not be accurately constrained.

\section{Conclusions}\label{sec:Conclusion}
 In this work,  we publish 56 new detached binaries selected from LAMOST MRS survey combined with 128 detached eclipsing binaries with independent atmospheric parameters (\teff, \logg, [M/H]) and accurate masses and radii compiled from previous studies. For the 56 LAMOST observed detached binaries, we perform the MCMC method with PHOEBE to obtain the orbital solutions.
With the additional LAMOST stars, we are able to increase the samples at each [M/H] bin. In particular, we extend the samples to lower [M/H] for high \teff\ stars. Hence, the distribution of [M/H] and \teff\ becomes more continuous and densified.
In total, we provide a catalog containing 128 binaries as the benchmark of stellar mass and radius covering a wide range of stellar parameters. The measurement uncertainties of masses and radii are within 5\%. 
 
We compared the samples with PARSEC model in different [M/H] and found that the observed data, including the new 56 LAMOST stars and the 128 stars from literature, essentially well match with the PARSEC isochrones.  In addition, we also find that the enlarged radii at around the turn-off point depend more on stellar ages than rotational velocity. Therefore, the radius estimates of the turn-off stars can also be potentially used as an indicator of age.

The comparisons between dynamical masses and radii with those derived from the stellar models show that the precision of model-estimated mass is $>$10\% and that of model-estimated radius is $>$15\% based on atmospheric parameters. 

\section*{}
This work is supported by the National Key R\&D Program of China No. 2019YFA0405500. C.L. Thanks the National Natural Science Foundation of China (NSFC) with grant Nos.11835057, 12173047, 12073047. The Guoshoujing Telescope (the Large Sky Area Multi-Object Fiber Spectroscopic Telescope LAMOST) is a National Major Scientific Project built by the Chinese Academy of Sciences. Funding for the project has been provided by the National Development and Reform Commission. LAMOST is operated and managed by the National Astronomical Observatories, Chinese Academy of Sciences.

\bibliography{sample631}{}
\bibliographystyle{aasjournal}

\end{document}